\documentclass[manuscript]{aastex}
\usepackage{apjfonts}
\usepackage{epsfig}
\usepackage{graphics}
\usepackage{natbib,graphicx}

\newcommand{\esinw}{\ensuremath{e \sin\omega}}
\newcommand{\ecosw}{\ensuremath{e \cos\omega}}
\newcommand{\teff}{\ensuremath{T_{\rm eff}}}

\newcommand{\logg}{\ensuremath{\log{g}}}

\newcommand{\feh}{[Fe/H]}
\newcommand{\rprs}{\ensuremath{R_{\rm p}/R_\star}}
\newcommand{\rhostar}{\ensuremath{\rho_\star}}

\begin{document}
\title{All Six Planets Known to Orbit Kepler-11 Have Low Densities}
\newcommand{\ikt}{{\it Kepler}}
\newcommand{\ik}{{\it Kepler~}}

\author{Jack J. Lissauer\altaffilmark{1}, Daniel Jontof-Hutter\altaffilmark{1}, Jason F. Rowe\altaffilmark{1,2}, Daniel C. Fabrycky\altaffilmark{3}, Eric D. Lopez\altaffilmark{4}, Eric Agol\altaffilmark{5}, Geoffrey W. Marcy\altaffilmark{6},  Katherine M. Deck\altaffilmark{7},  Debra A. Fischer\altaffilmark{8},  Jonathan J. Fortney\altaffilmark{4}, Steve B. Howell\altaffilmark{1}, Howard Isaacson\altaffilmark{6},  Jon M. Jenkins\altaffilmark{1,2}, Rea Kolbl\altaffilmark{6}, Dimitar Sasselov\altaffilmark{9}, Donald R. Short\altaffilmark{10}, William F. Welsh\altaffilmark{11}}

\email{ Jack.Lissauer@nasa.gov }
\altaffiltext{1}{NASA Ames Research Center, Moffett Field, CA 94035, USA}
\altaffiltext{2}{SETI Institute/NASA Ames Research Center, Moffett Field, CA 94035, USA}
\altaffiltext{3}{Department of Astronomy and Astrophysics, University of Chicago, 5640 South Ellis Avenue, Chicago, IL 60637, USA}
\altaffiltext{4}{Department of Astronomy \& Astrophysics, University of California, Santa Cruz, CA 95064, USA}
\altaffiltext{5}{Department of Astronomy, Box 351580, University of Washington, Seattle, 
WA 98195, USA}
\altaffiltext{6}{Astronomy Department, University of California, Berkeley, CA 94720, USA} 
\altaffiltext{7}{Department of Physics and Kavli Institute for Astrophysics and Space Research, Massachusetts Institute of Technology, 77 Massachusetts Ave., Cambridge, MA 02139, USA}
\altaffiltext{8}{Department of Astronomy, Yale University, New Haven, CT 06520-8101, USA}
\altaffiltext{9}{Harvard-Smithsonian Center for Astrophysics, 60 Garden Street, Cambridge, MA 02138, USA}
\altaffiltext{10}{Department of Mathematics, San Diego State University, 5500 Campanile Drive, San Diego, CA 92182, USA}
\altaffiltext{11}{Astronomy Department, San Diego State University, 5500 Campanile Drive, San Diego, CA 92182, USA}

\begin{abstract}
The Kepler-11 planetary system contains six transiting planets ranging in size from 1.8 to 4.2 times the radius of Earth. Five of these planets orbit in a tightly-packed configuration with periods between 10 and 47 days.  We perform a dynamical analysis of the system based upon transit timing variations observed in more than three years of \ik photometric data.  Stellar parameters are derived using a combination of spectral classification and constraints on the star's density derived from transit profiles together with planetary eccentricity vectors provided by our dynamical study.  Combining masses of the planets relative to the star from our dynamical study and radii of the planets relative to the star from transit depths together with deduced stellar properties yields  measurements of the radii of all six planets, masses   of the five inner planets, and an upper bound to the mass of the outermost planet, whose orbital period is 118 days.  We find mass-radius combinations for all six planets that imply that substantial fractions of their volumes are occupied by constituents that are less dense than rock. Moreover, we examine the stability of these envelopes against photo-evaporation and find that the compositions of at least the inner two planets have likely been significantly sculpted by mass loss. The Kepler-11 system contains the lowest mass exoplanets for which both mass and radius have been  measured.
\end{abstract}
\section{Introduction}
Within our Solar System, Earth and smaller bodies are primarily rocky (or, far from the Sun, mixtures of rock and ices), whereas the cosmically-abundant low-density constituents H$_2$ and He dominate the volume in Uranus/Neptune and larger bodies.  There are no local examples of bodies intermediate in size or mass between Earth (1 R$_\oplus$, 1 M$_\oplus$) and  Uranus/Neptune, both of which are larger than 3.8 R$_\oplus$ and more massive than 14 M$_\oplus$. However, observations of extrasolar planets are now filling this gap in our knowledge of the mass-radius relationship of planetary bodies.

To date, the only accurate radius measurements for exoplanets have been provided by planets observed to transit across the disk of their star.  The fractional depth of the transit provides a direct measure for the ratio of the radius of the planet to that of its star. The star's radius is estimated using spectroscopic classification, in some cases augmented by other techniques. Doppler measurements of the variation of a star's radial velocity have been used to compute mass estimates for almost two hundred transiting giant planets, as well as for the first three sub-Uranus exoplanets for which both radii and masses were determined: GJ 1214 b \citep{cha09}, CoRoT-7 b \citep{que09}, and Kepler-10 b \citep{bat11}. Analysis of transit timing variations (TTVs) resulting from mutual planetary perturbations provided dynamical estimates of the masses of the five innermost known planets orbiting Kepler-11 \citep{liss11a}, more than doubling the number of exoplanets less massive than Uranus with both size and mass measurements. Precise mass estimates have subsequently been obtained for several more sub-Uranus mass planets, in three cases by using radial velocity (RV): 55 Cancre e (\citealt{win11,endl12}), Kepler-20 b \citep{gau12}, and GJ 3470 b \citep{bon12};  three using  TTVs: Kepler-36 b,c \citep{car12}, and Kepler-30 b \citep{san12}; and one, Kepler-18 b \citep{coch11}, using a combination of RV and TTV data. Less precise estimates for the masses of dozens of \ik planets and planet candidates, many of which are in this mass range, have been derived from TTVs by \citet{wu12}.

\citet{liss11a} estimated the masses of the five planets Kepler-11 b-f using only the first 16 months of \ik data. Similar mass constraints on these planets, as well as an upper limit of 30 M$_\oplus$ on the mass of the outer planet Kepler-11 g, were obtained by \citet{mig12}. \citet{mig12} analyzed the same Q1-Q6\footnote{The 
\ik spacecraft rotates four times per orbit to keep the sunshade and solar panels oriented properly. Targets are imaged on different parts of the 
focal plane during different orientations.  The \ik orbital period is $\sim$372 days, and the data are grouped according to the ``quarter'' year during 
which observations were made.  The data on Kepler-11 taken prior to 
\ik's first ``roll'' are referred to as Q1.  Subsequent quarters are 
numbered sequentially: Q2, Q3, ...} data 
using a photodynamical model, which adjusted planetary parameters (size, orbital elements, masses) to minimize the residuals of a fit of a model lightcurve that accounts for mutual planetary interactions to the measured lightcurve. 

We report herein more precise estimates of the masses of the six Kepler-11 planets derived from TTV measurements that incorporate 40 months of \ik photometric time series data.  In Section 2, we present our estimates of transit times; detailed descriptions of the three independent techniques used to compute these times are given in Appendix A.  Our dynamical analysis of the Kepler-11 system based upon these transit times is presented in Section 3, with additional information provided in Appendix B.  In Section 4, we combine estimates of stellar density obtained using transit profiles and the dynamical measurement of planetary eccentricities presented in Section 3 together with  analyses of high-resolution spectra taken at the Keck I telescope to provide refined parameters for the star Kepler-11.  We tabulate the properties of Kepler-11's six known planets that are derived by combining lightcurve analysis with our dynamical results and stellar parameters in Section 5, wherein we also discuss implications of these results for planetary compositions.  We conclude the paper with a summary of our principal results.

\section{Measurement of Transit Times from \ik Photometric Time Series}

Variations in the brightness of Kepler-11 have been monitored with an effective duty cycle exceding 90\%   starting at barycentric Julian date (BJD) 2454964.512, 
 with all data returned to Earth at a cadence of 29.426 minutes (long cadence, LC); data have also been returned  at a cadence of 58.85 seconds (short cadence, SC) since BJD 2455093.216. 
 Our analysis uses short cadence data where available, augmented by the long cadence dataset primarily during the epoch prior to  BJD 2455093.216, for which no SC data were returned to Earth. We obtained these data from the publicly-accessible MAST archive at http://archive.stsci.edu/kepler/ . 

As measurement of transit times (TTs) requires a complicated analysis of often noisy data, authors Jason Rowe (J.R.), Eric Agol (E.A.) and Donald Short (D.S.) performed independent measurements of TTs using techniques described in Appendix A. Figure~\ref{fig:1} shows the deviations of all three sets of observed transit times, $O$, relative to time from a linear ephemeris fit, $C_\textit{l}$, through Q14 \ik data. Here and throughout we base our timeline for transit data from JD-2,454,900.  

As evident in Figure~\ref{fig:1}, each set of TT measurements contain several outliers. These outliers  are unlikely
to be correct, and may be due to overlapping transits,
star spots, or uncertain fits to the lightcurve. Trying to fit these outlier TTs would degrade our dynamical studies. Therefore, we remove points where
only one of the methods yields a TT whose uncertainty is more than 2.5 times as large as the median TT uncertainty computed by that method for the planet in question.
We then use the three sets of measured TTs to filter out unreliable measurements as follows:  If two or three sets of measurements are available for a specific transit and each of the $1\sigma$ uncertainty ranges overlap with at least one of the other ranges, then each of the points are used. If there is only a single measurement, or if there is no overlap of $1\sigma$ uncertainty ranges, then all measurements of this transit are discarded. If three measurements are available, and two overlap but the third does not overlap with either,  then the data are discarded for TTs of planets b -- f, but the two overlapping points are retained for planet g, which has far fewer transits observed than any other planet (and no significant TTVs even with these points included). This culling procedure removed  fewer than 8\% of detected transit times from  each dataset, with the most points discarded from Kepler-11 b, whose transits are the most numerous and have the lowest signal-to-noise ratio (S/N). For planet b, we removed 17 of the 103 TTs measured by E.A., 9 of the 111 TTs measured by J.R., and 13 of the 90 TTs measured by D.S. 
 Our approach is conservative in the sense that the data set used for our dynamical studies presented in Section 3 consists only of transit times that are corroborated by at least one alternative method.

\begin{figure}
\includegraphics [height = 1.8 in]{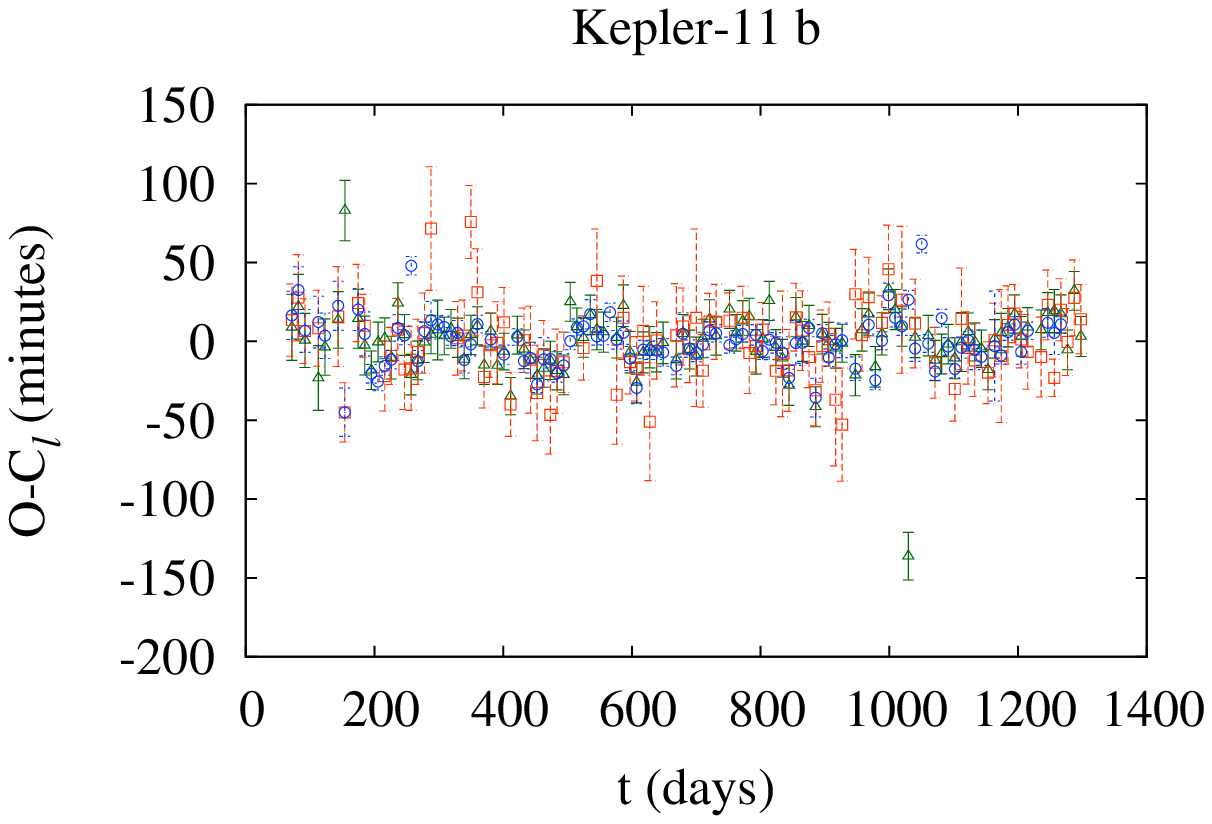}
\includegraphics [height = 1.8 in]{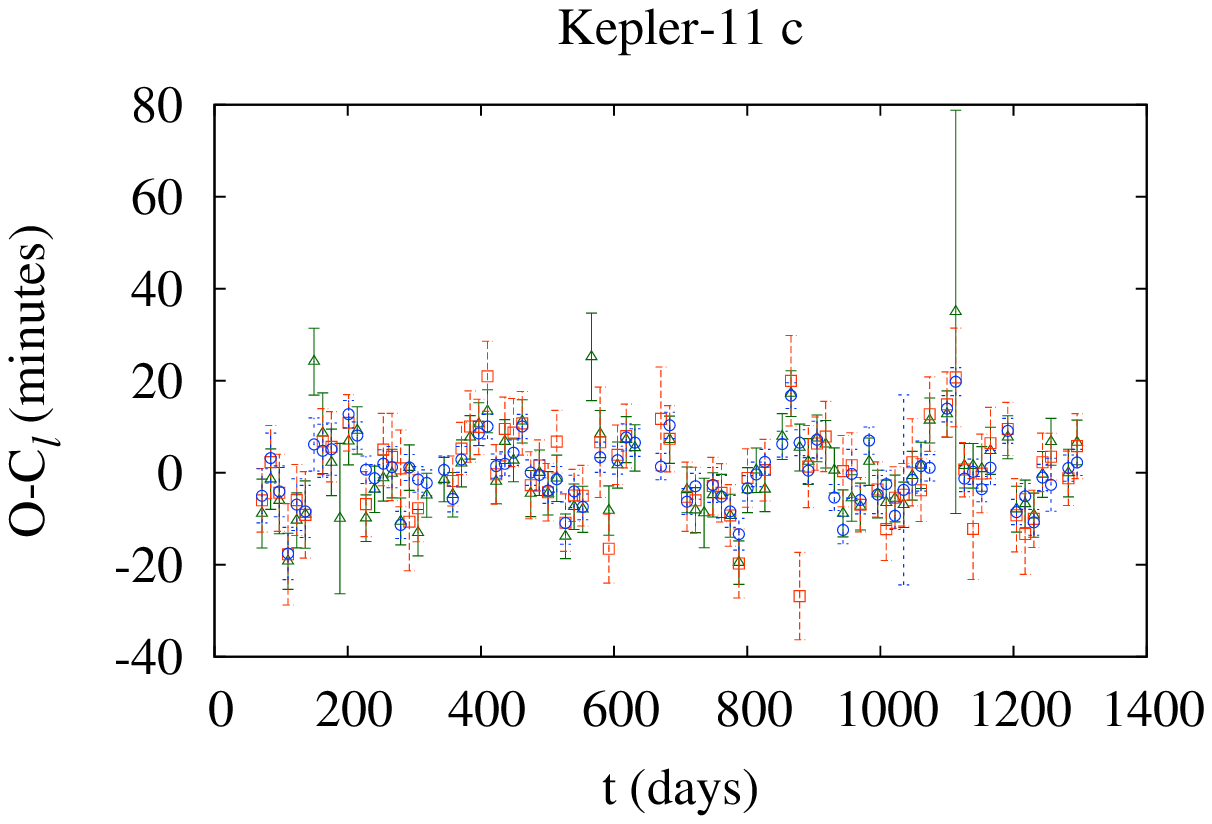}
\newline
\includegraphics [height = 1.8 in]{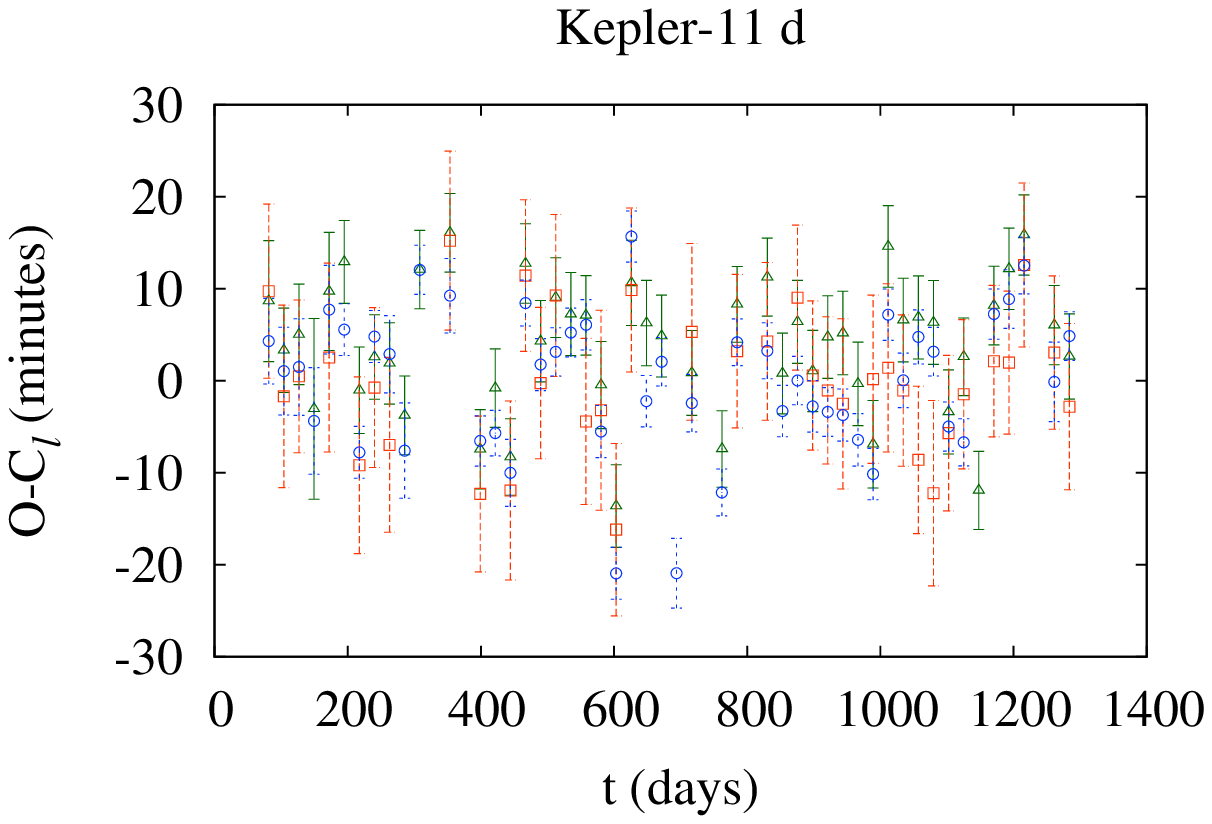}
\includegraphics [height = 1.8 in]{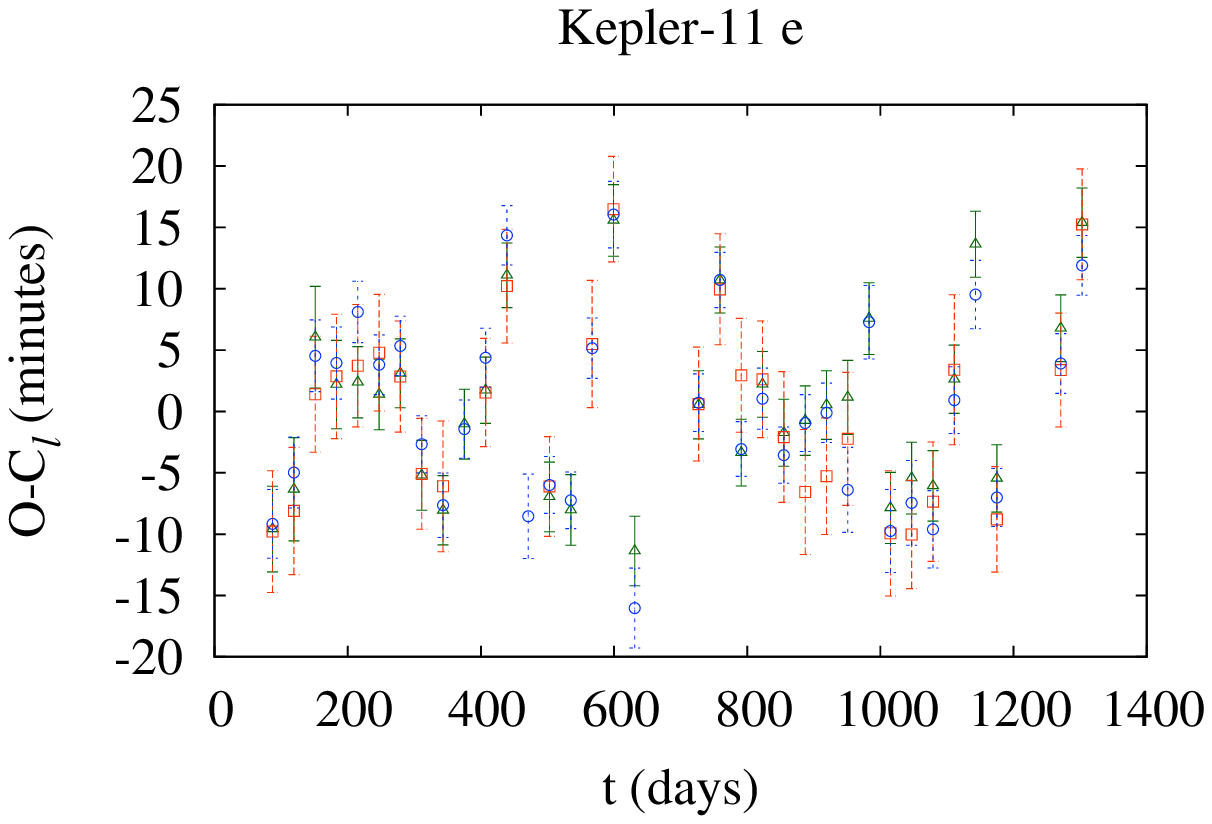}
\newline
\includegraphics [height = 1.8 in]{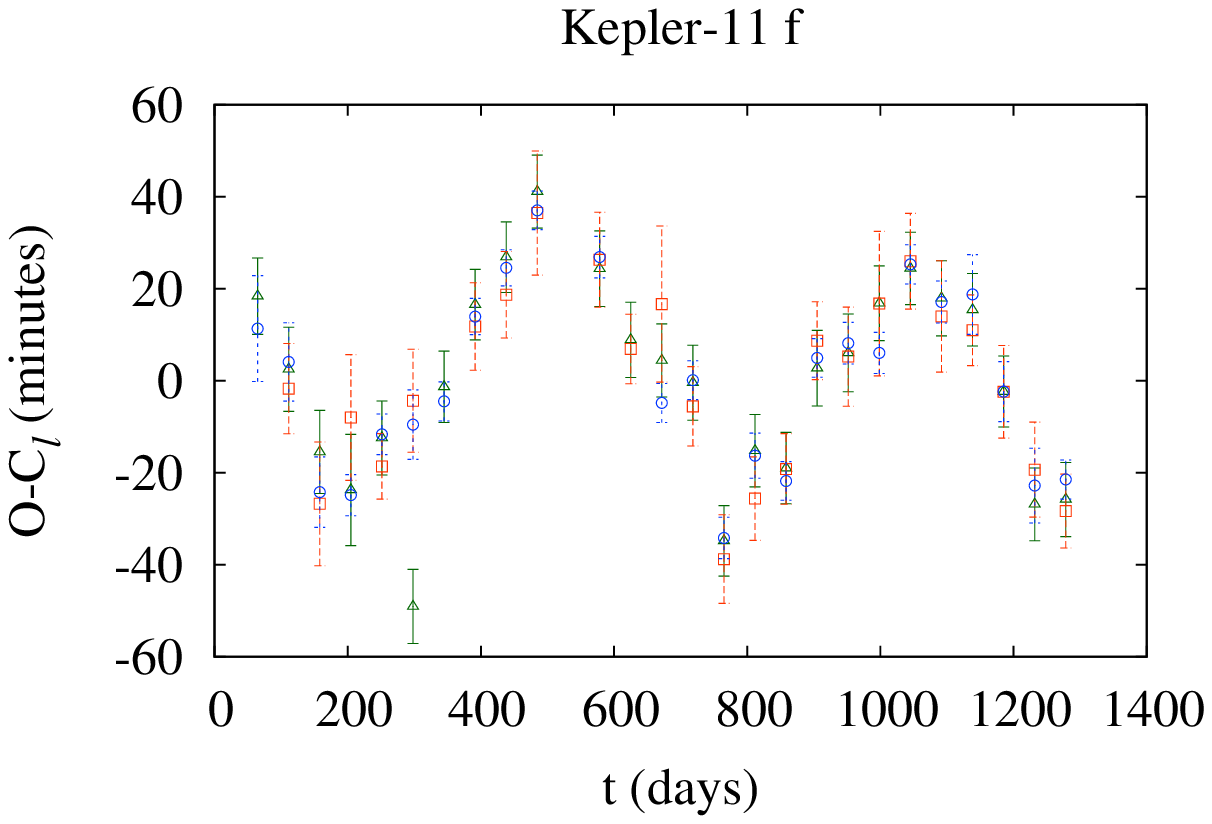}
\includegraphics [height = 1.8 in]{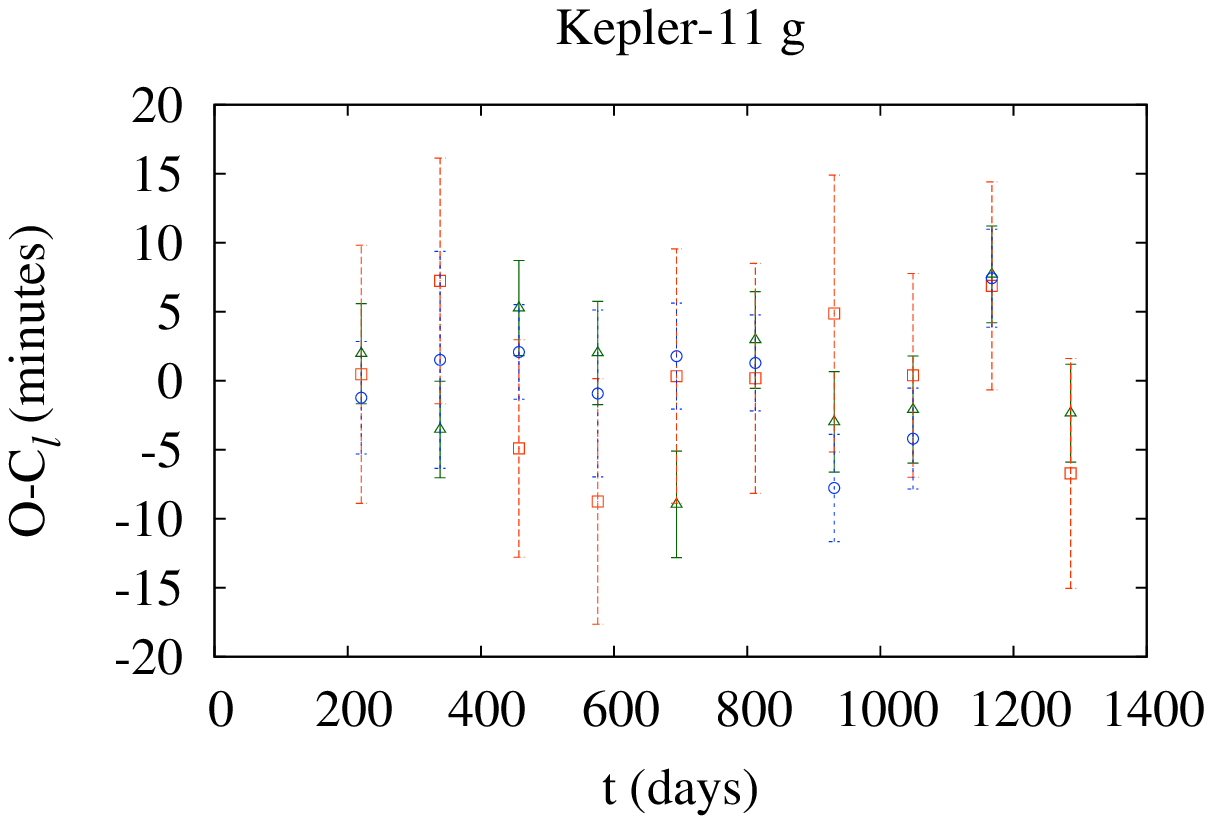}
\caption{Transit timing variations for Kepler-11's six known planets, using short cadence data when available, supplemented by long cadence data prior to $t$ (JD-2,454,900) = 193 days, where short cadence data were not sent to Earth. The TTs measured by E.A. are displayed as green open triangles, those from J.R. as blue open circles, and those calculated by D.S. as red open squares, with their respective methods described in Appendix A. The sets of data points are largely consistent. The observed transit times, $O$, are displayed as deviates from times, $C_{\textit{l}}$, that were calculated using a linear fit to each set of transit data, i.e., a fit that assumes strictly periodic orbits. All measured TTs are displayed, apart from one outlier for Kepler-11 d that deviated from both of the other estimates and a linear ephemeris by more than three hours. Note that the vertical scales differ among panels.}
\label{fig:1} 
\end{figure}

\section{Dynamical Models of the Kepler-11 Planetary System}

Transits of a planet on a keplerian orbit about its star must be strictly periodic.  In contrast, the gravitational interactions among planets in a multiple planet system cause orbits to speed up and slow down by small amounts, leading to deviations from exact periodicity of transits (\citealt{dob96,hm05,assc05}). Such variations are strongest when planetary orbital periods are commensurate or nearly so, which is the case for the large planets Kepler-9 b and c \citep{hol10}, or when planets orbit close to one another, which is the case for the inner five transiting planets of Kepler-11 \citep{liss11a}.

To integrate planetary motions, we adopt the 8th order Runge-Kutta Prince-Dormand method, which has 9th order errors. Our choice of dynamical epoch was $T_{0} = 680$ days, near the midpoint of the fourteen quarters of \textit{Kepler} data being modeled. In all of our simulations, the orbital period and phase of each planet are free parameters. The phase is specified by the midpoint of the first transit subsequent to our chosen epoch. Initially, we keep all planetary masses as free parameters. In some cases, we required planets to be on circular orbits at epoch, whereas in others we allowed the orbits to be eccentric. 
 
 We have assumed co-planarity, i.e., negligible mutual inclinations between planetary orbits,  in all of our dynamical models. We make no attempt to model transit durations or impact parameters in our dynamical simulations. 

Our integrations produce an ephemeris of simulated transit times, $C_\textit{s}$, and we compare these simulated times to the observed TTs. We employ the Levenberg-Marquardt algorithm to search for a local minimum in $\chi^2$. The algorithm evaluates the local slope and curvature of the $\chi^2$ surface. Once it obtains a minimum, the curvature of the surface is used to evaluate error bars. Other parameters are allowed to float when determining the error limits on an individual parameter's error bars. Assuming that the $\chi^2$ surface is parabolic in the vicinity of its local minimum, its contours are concentric ellipses centered at the best-fit value. The orientations of these ellipses depend on correlations between parameters. The errors that we quote account for the increase in uncertainty in some dimensions due to such correlations.

\begin{figure}
\includegraphics [height = 2.1 in]{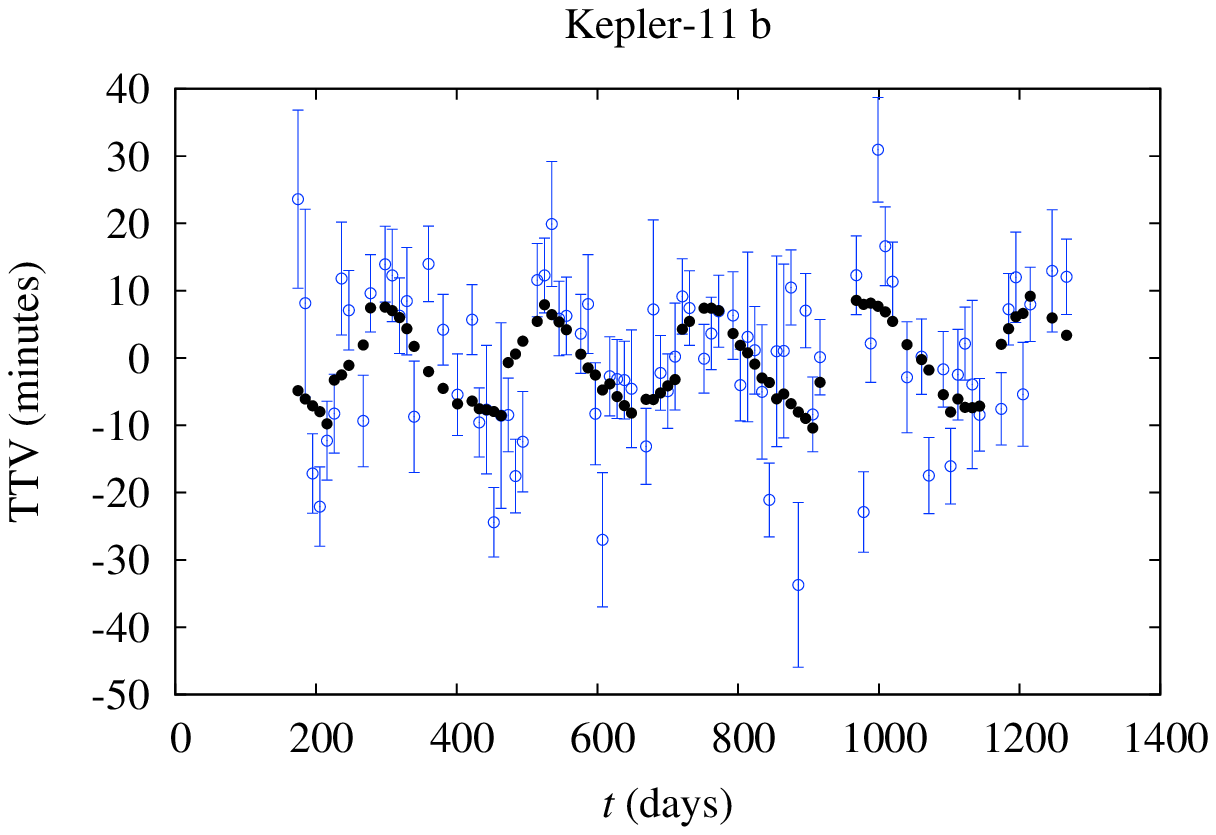}
\includegraphics [height = 2.1 in]{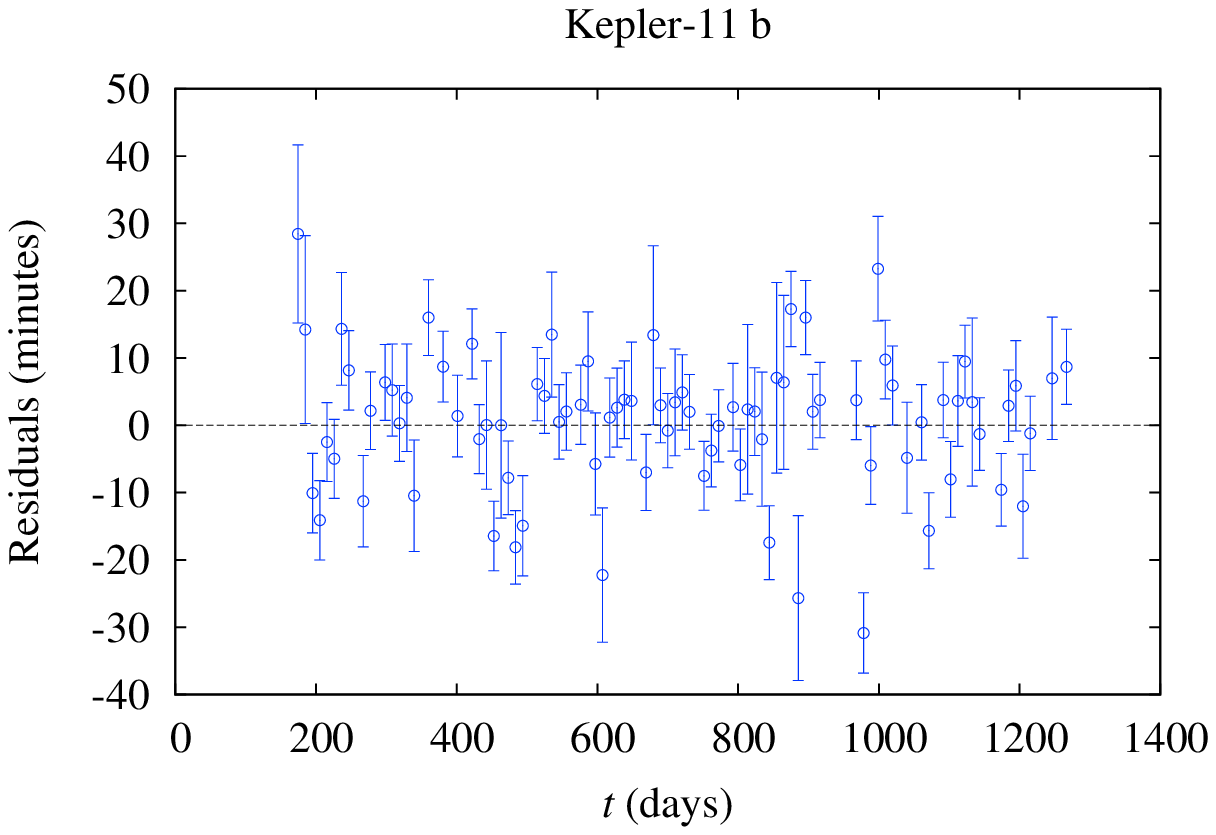}
\newline
\includegraphics [height = 2.1 in]{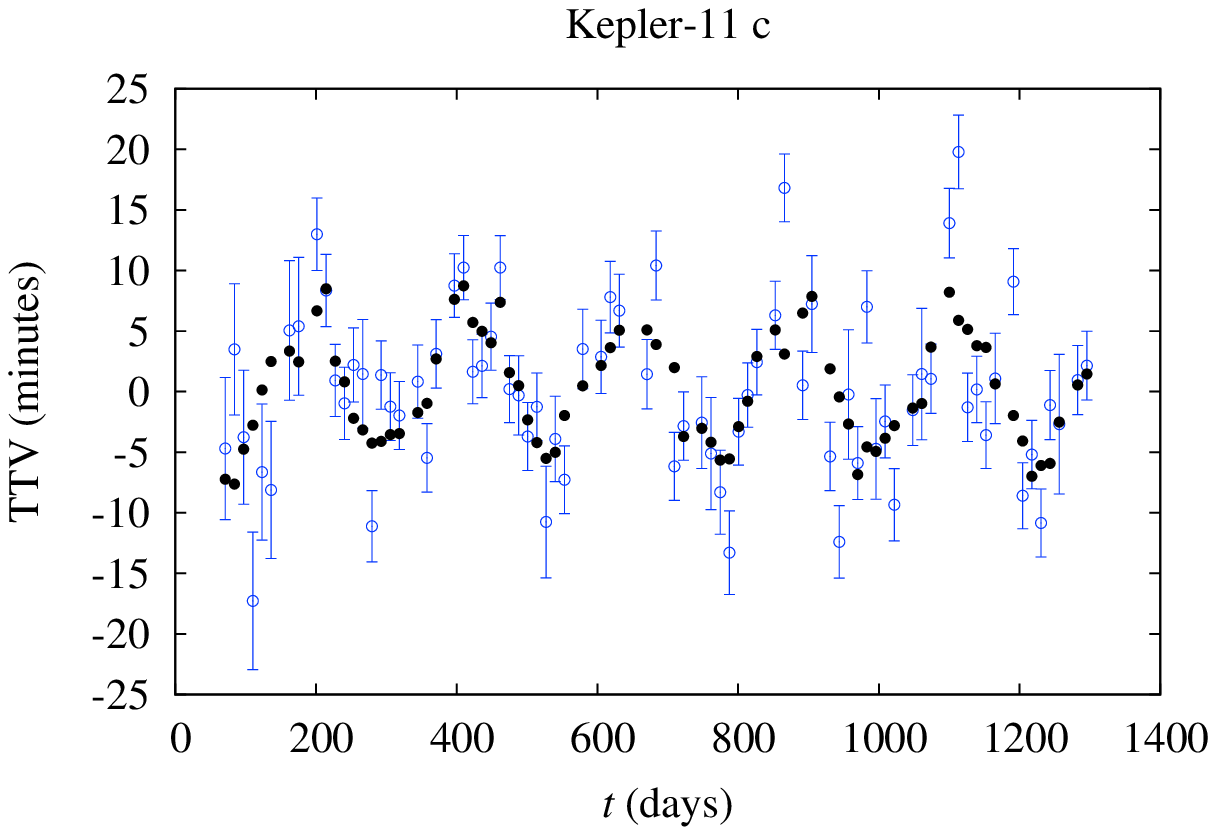}
\includegraphics [height = 2.1 in]{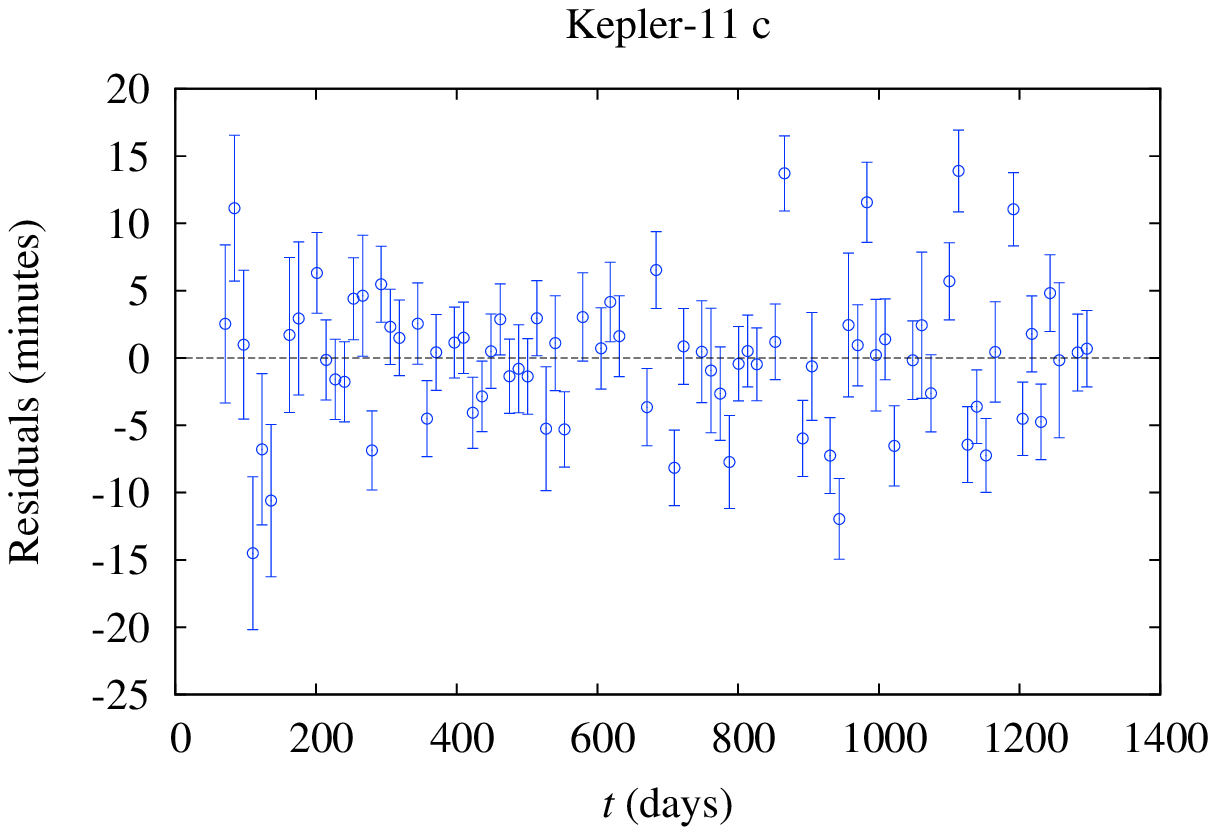}
\newline
\includegraphics [height = 2.1 in]{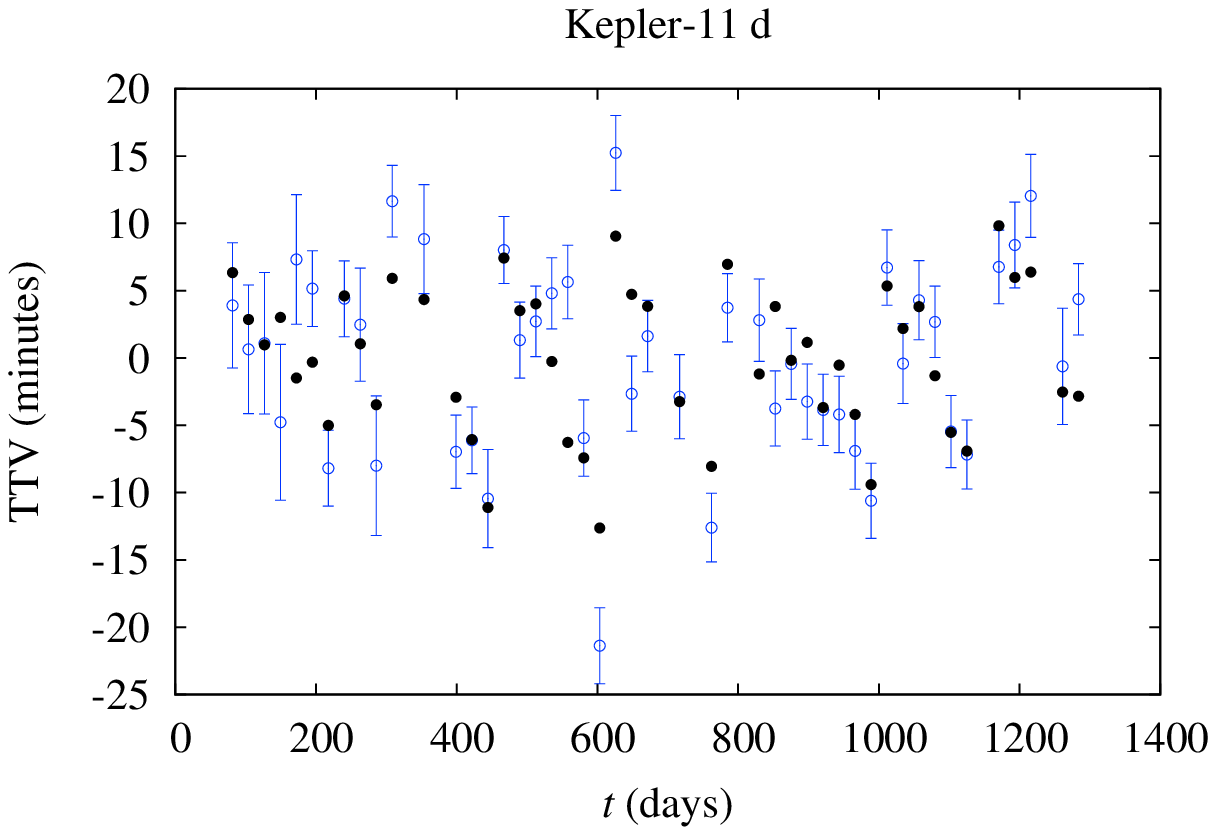}
\includegraphics [height = 2.1 in]{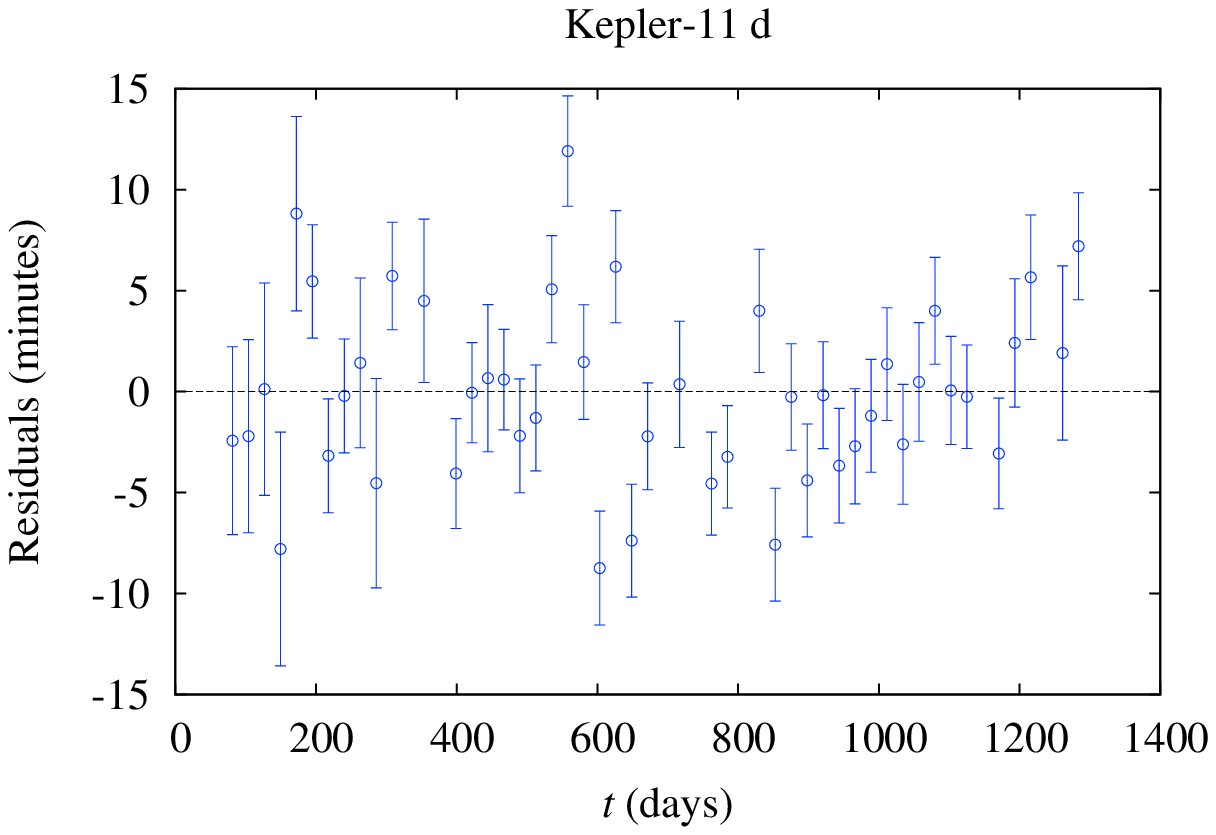}
\caption{Observed and simulated transit timing variations for planets Kepler-11 b, c and d, using transit measurements from E.A. The panels on the left-hand side compare observed TTVs (the difference between observed TTs and the best fit constant-period ephemeris, $O-C_{\textit{l}}$), which are represented by open symbols with error bars, with model TTVs (the departure of model times from the same constant-period ephemeris, $C_{\textit{s}} - C_\textit{l}$), which are represented by filled black points. The right hand side plots the residuals of the fit (i.e., the dynamical model subtracted from the observed transit times). Note the differences between the vertical scales of the various panels.
}
\label{fig:EA1} 
\end{figure}

\begin{figure}
\includegraphics [height = 2.1 in]{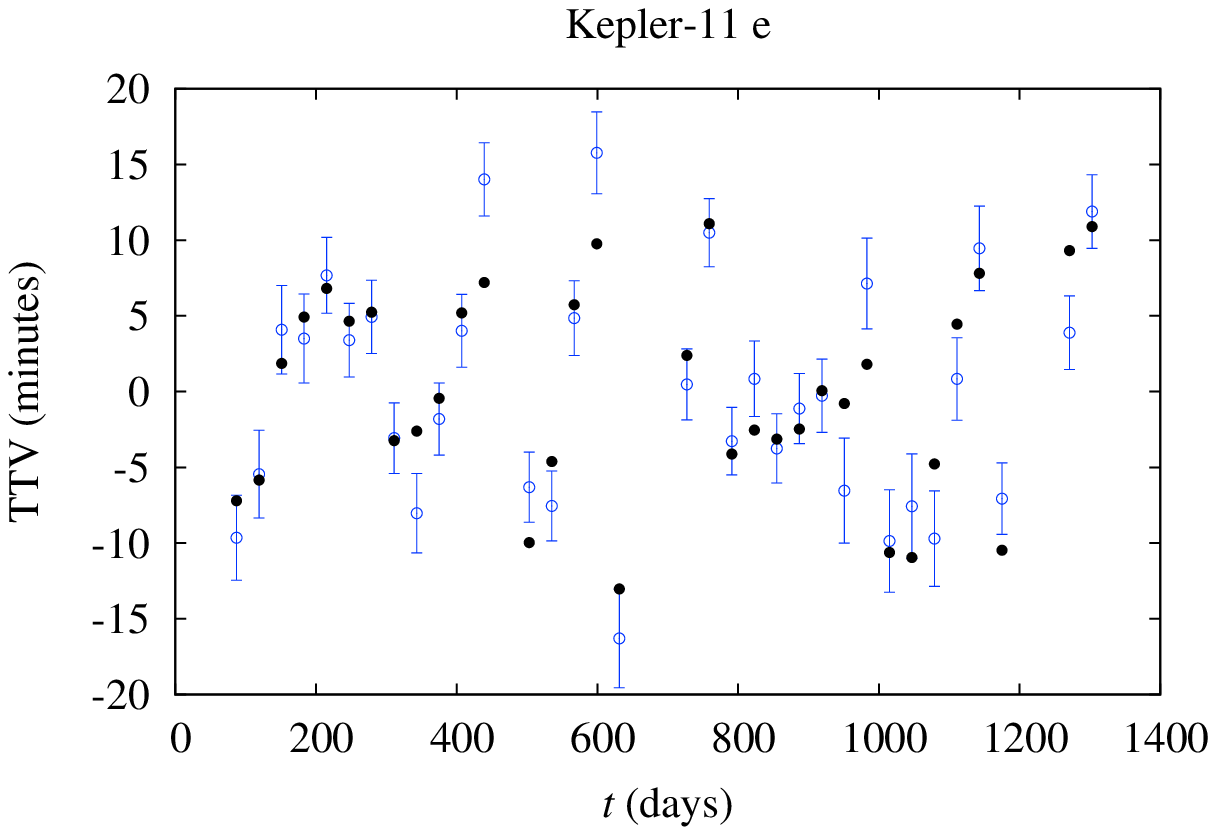}
\includegraphics [height = 2.1 in]{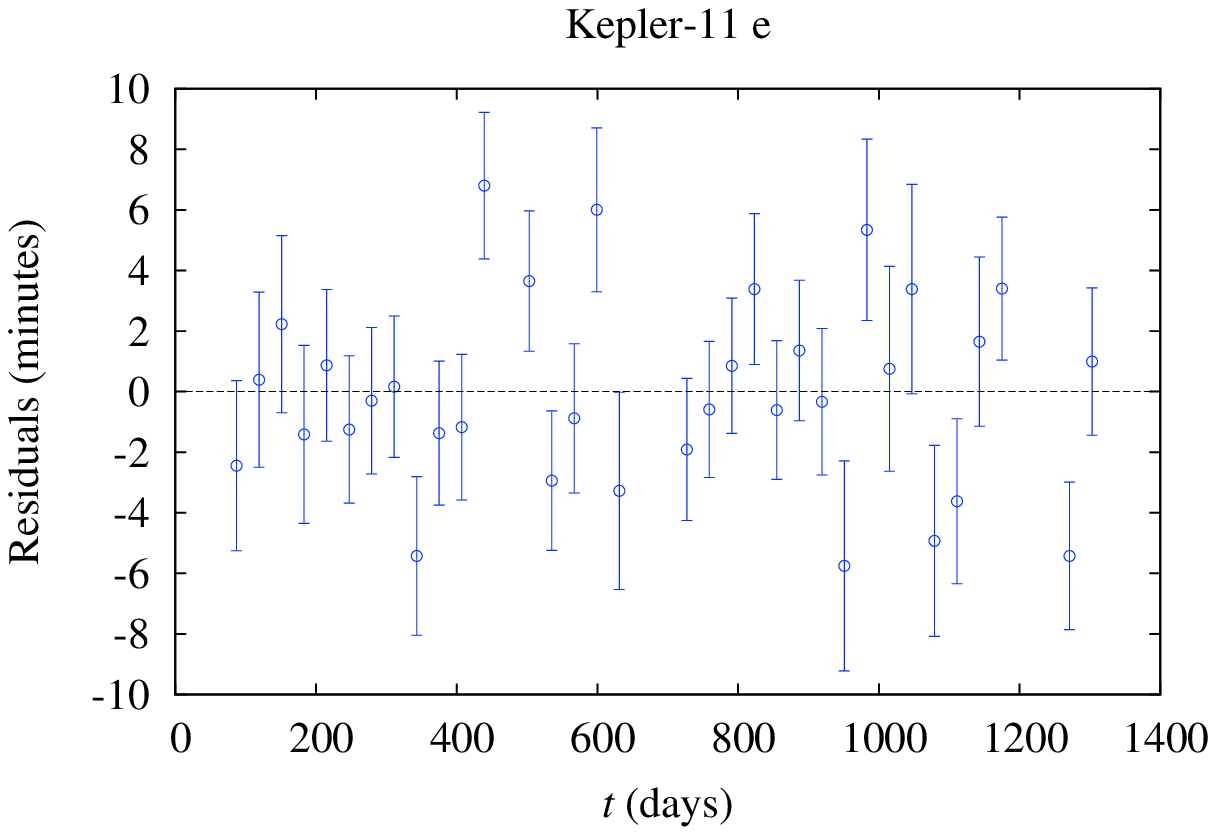}
\newline
\includegraphics [height = 2.1 in]{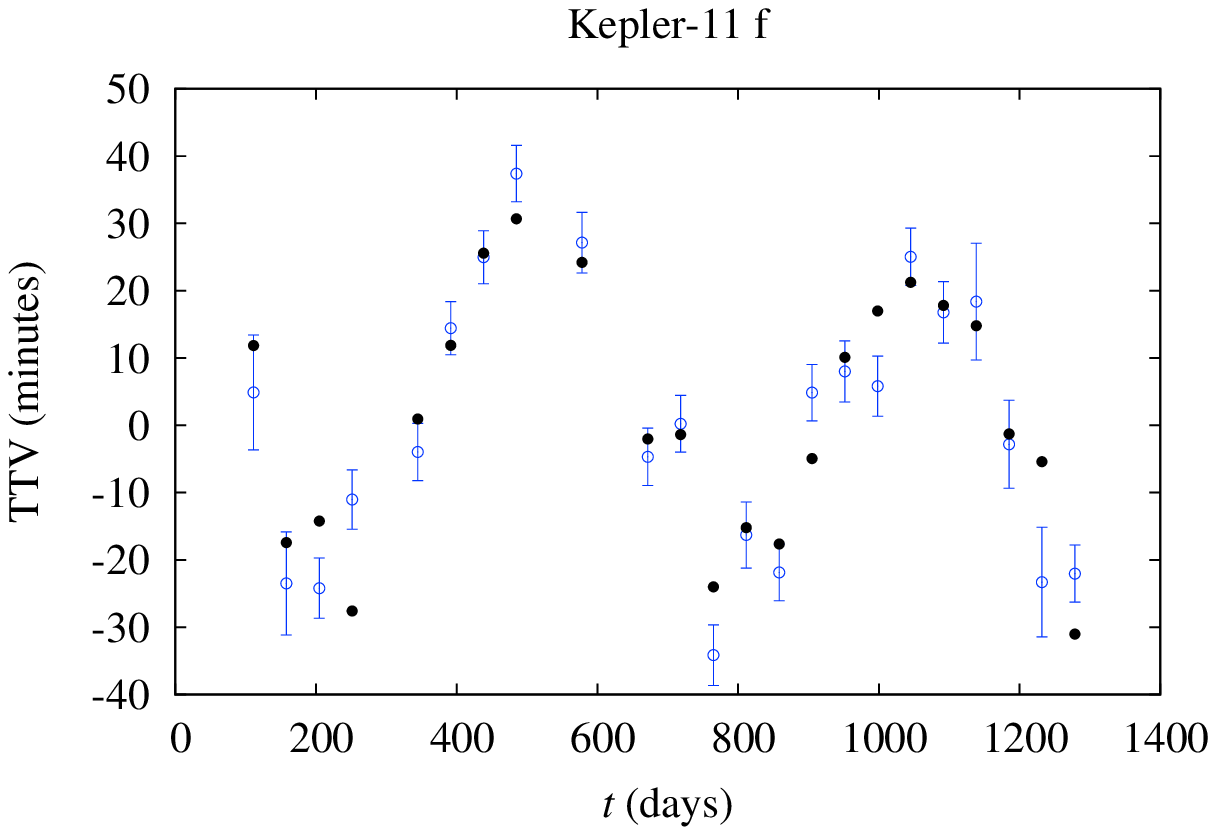}
\includegraphics [height = 2.1 in]{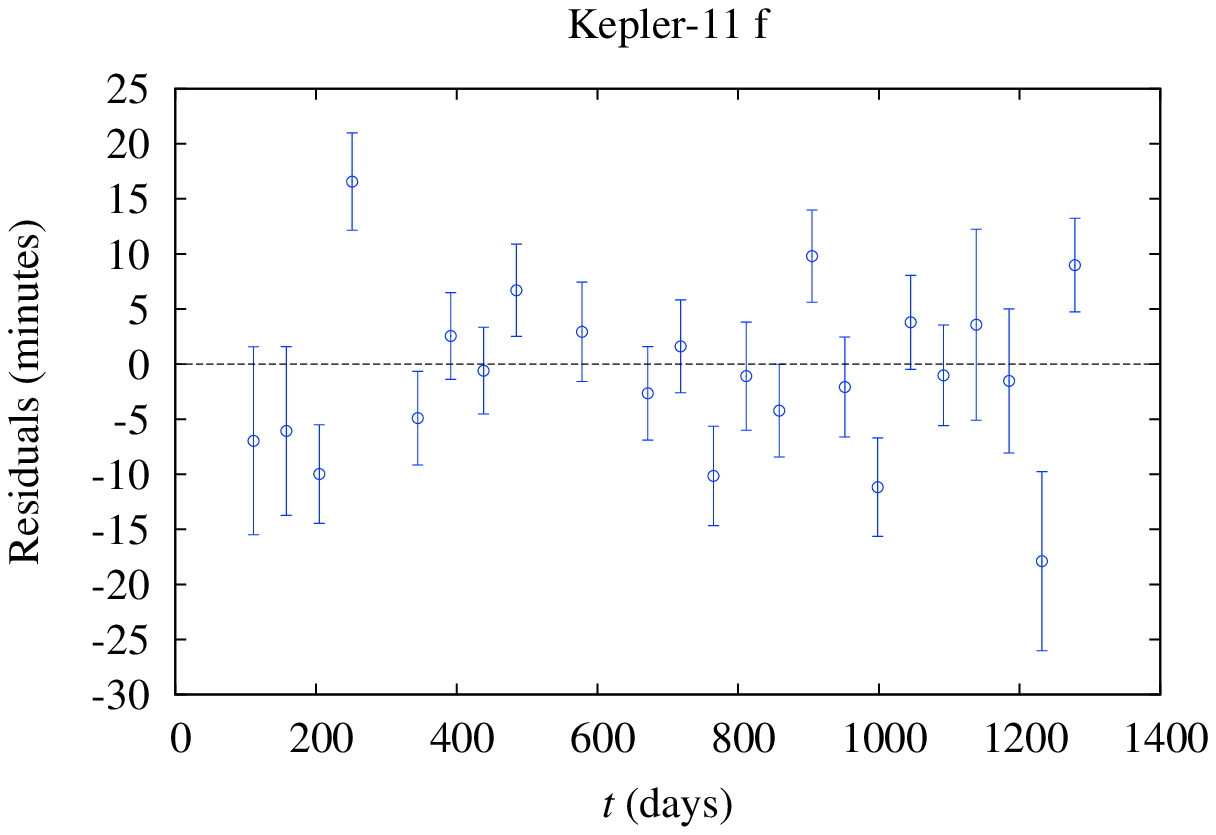}
\newline
\includegraphics [height = 2.1 in]{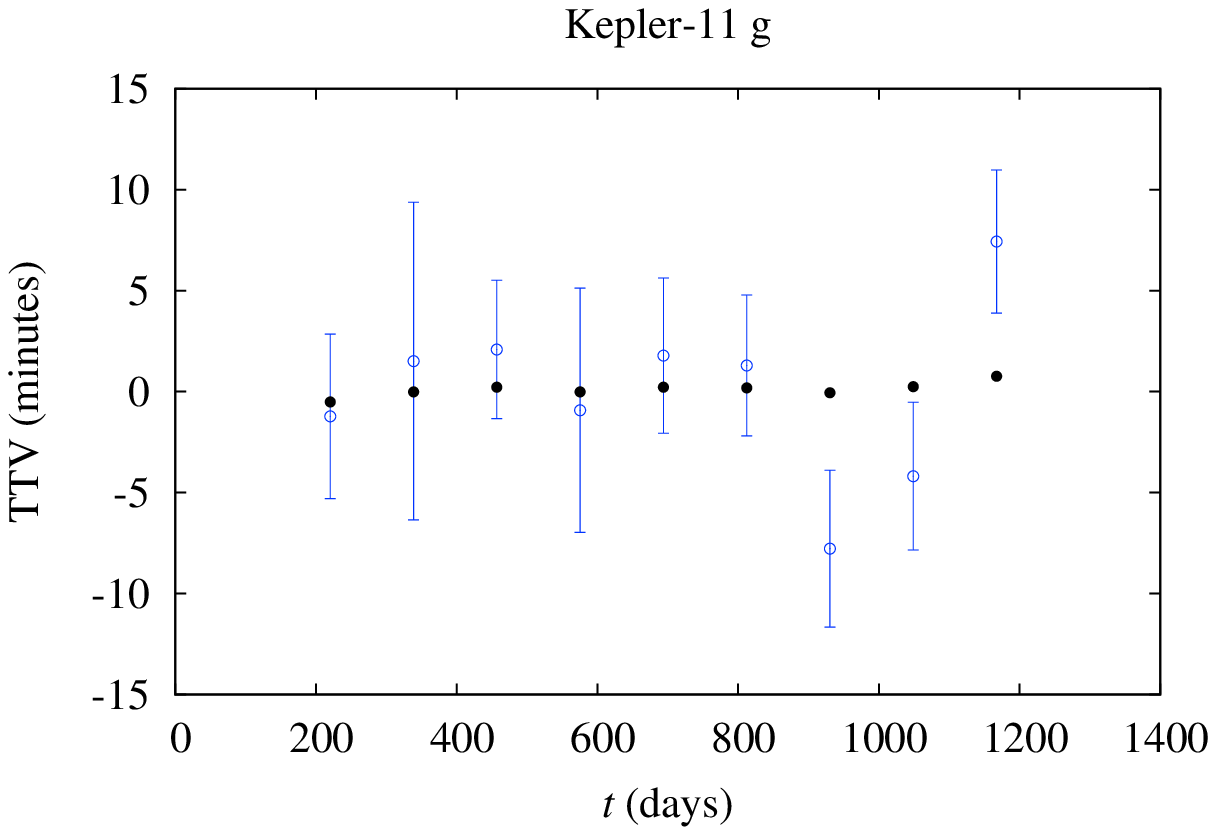}
\includegraphics [height = 2.1 in]{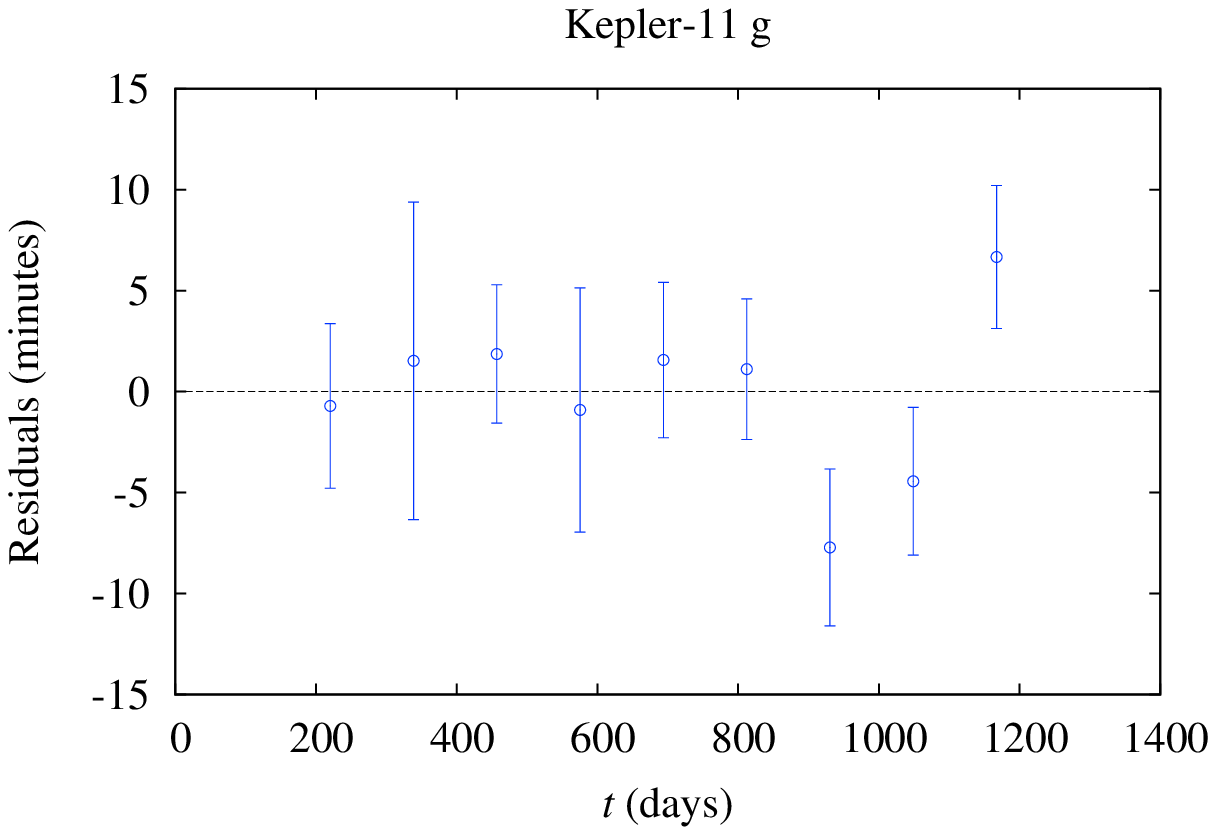}
\caption{Observed and simulated transit timing variations for Kepler-11 e, f and g, using transit time measurements from E.A.  See the caption to Figure 2 for details.}
\label{fig:EA2} 
\end{figure}

\begin{figure}
\includegraphics [height = 2.1 in]{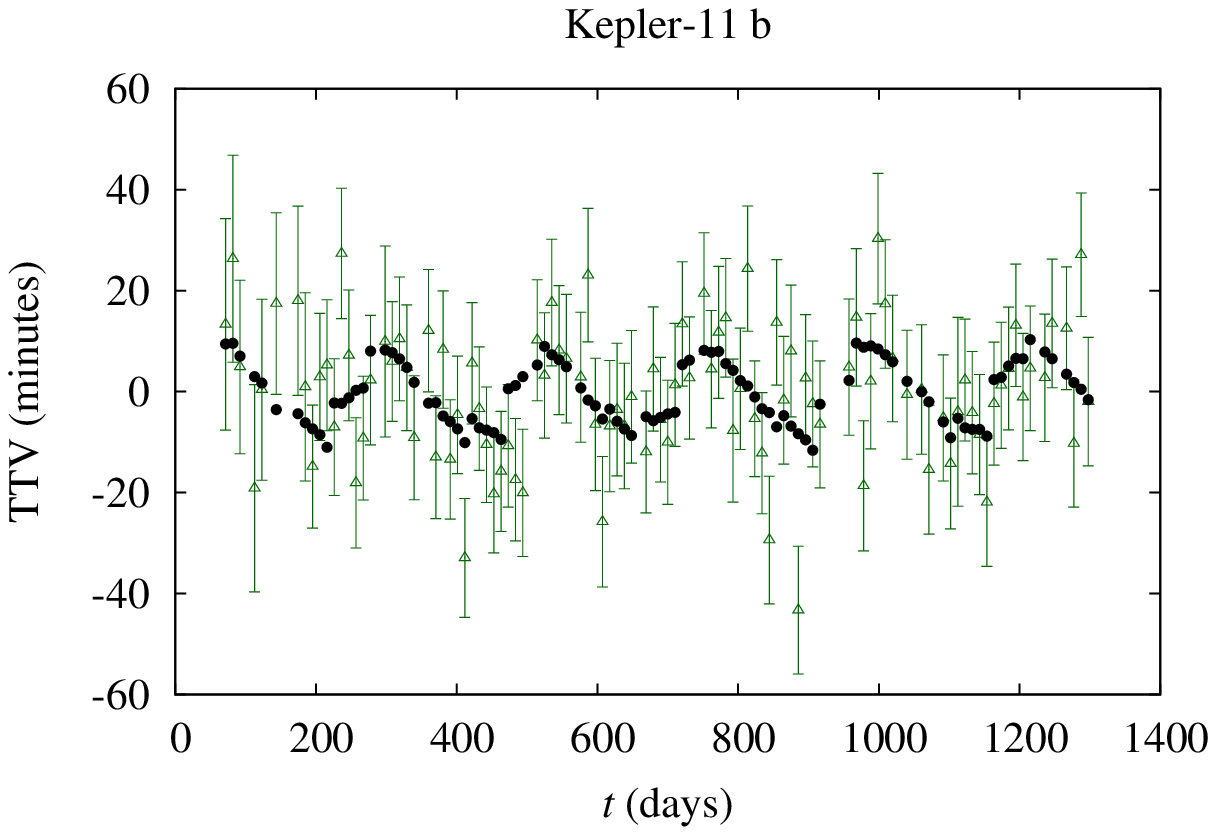}
\includegraphics [height = 2.1 in]{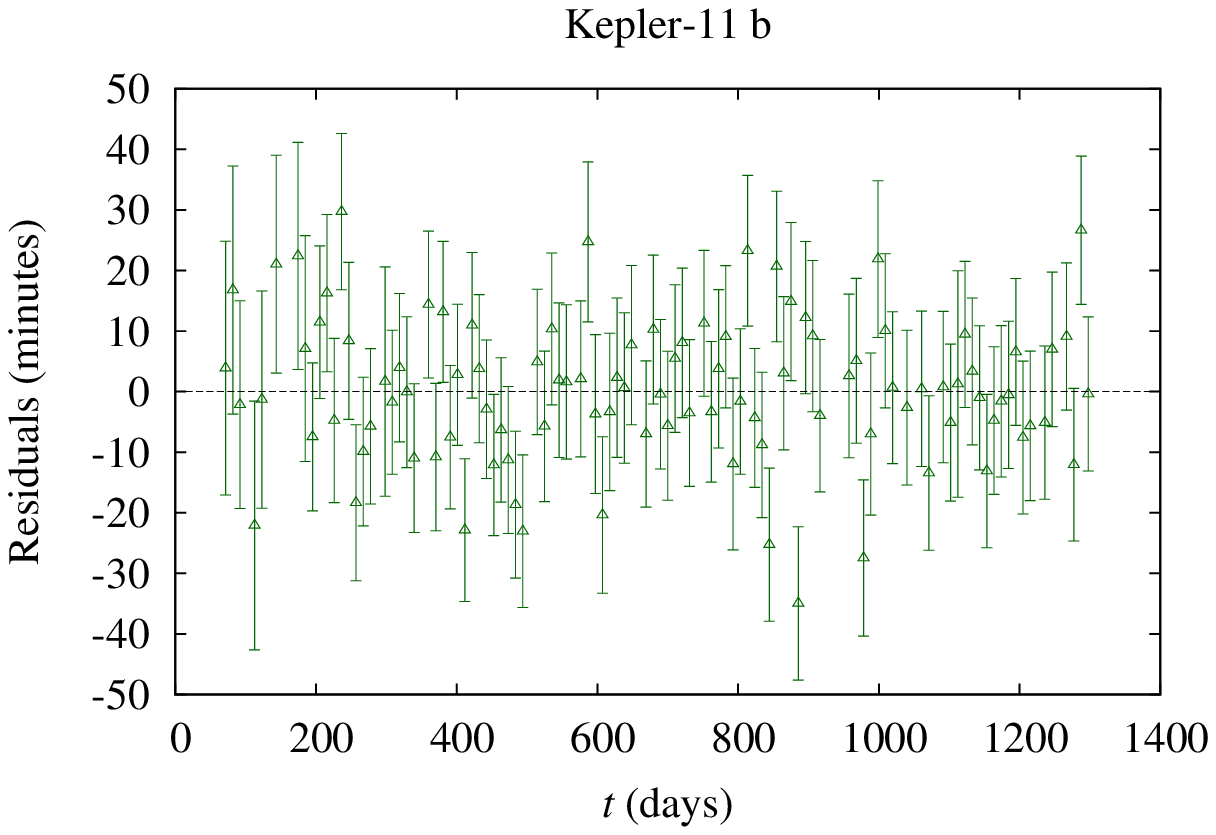}
\newline
\includegraphics [height = 2.1 in]{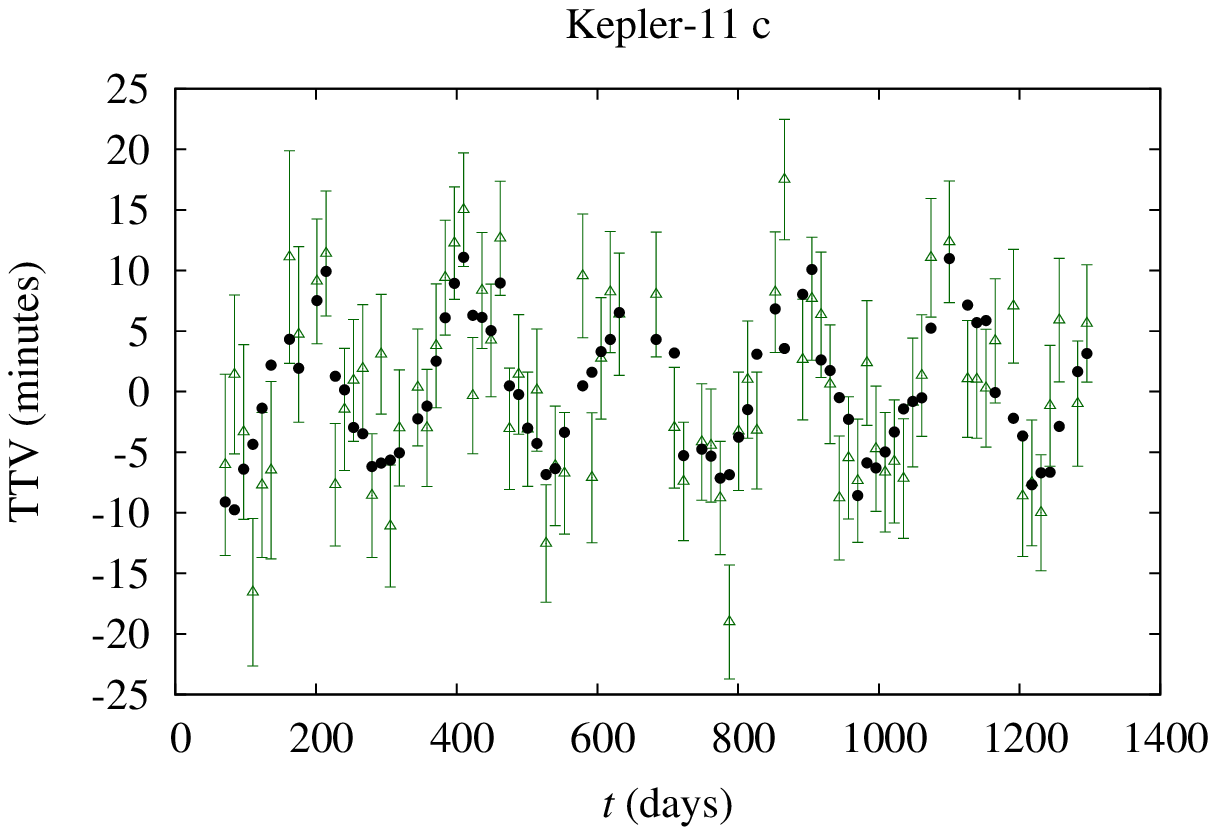}
\includegraphics [height = 2.1 in]{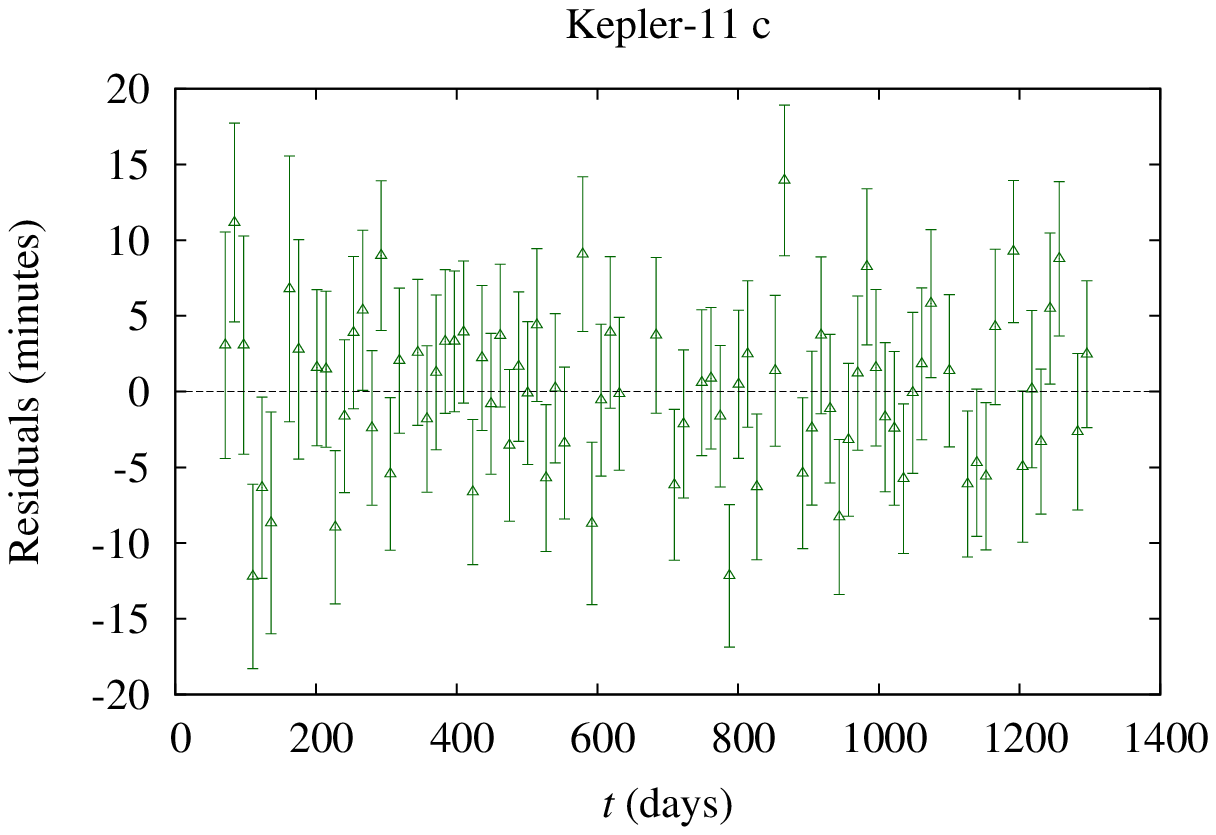}
\newline
\includegraphics [height = 2.1 in]{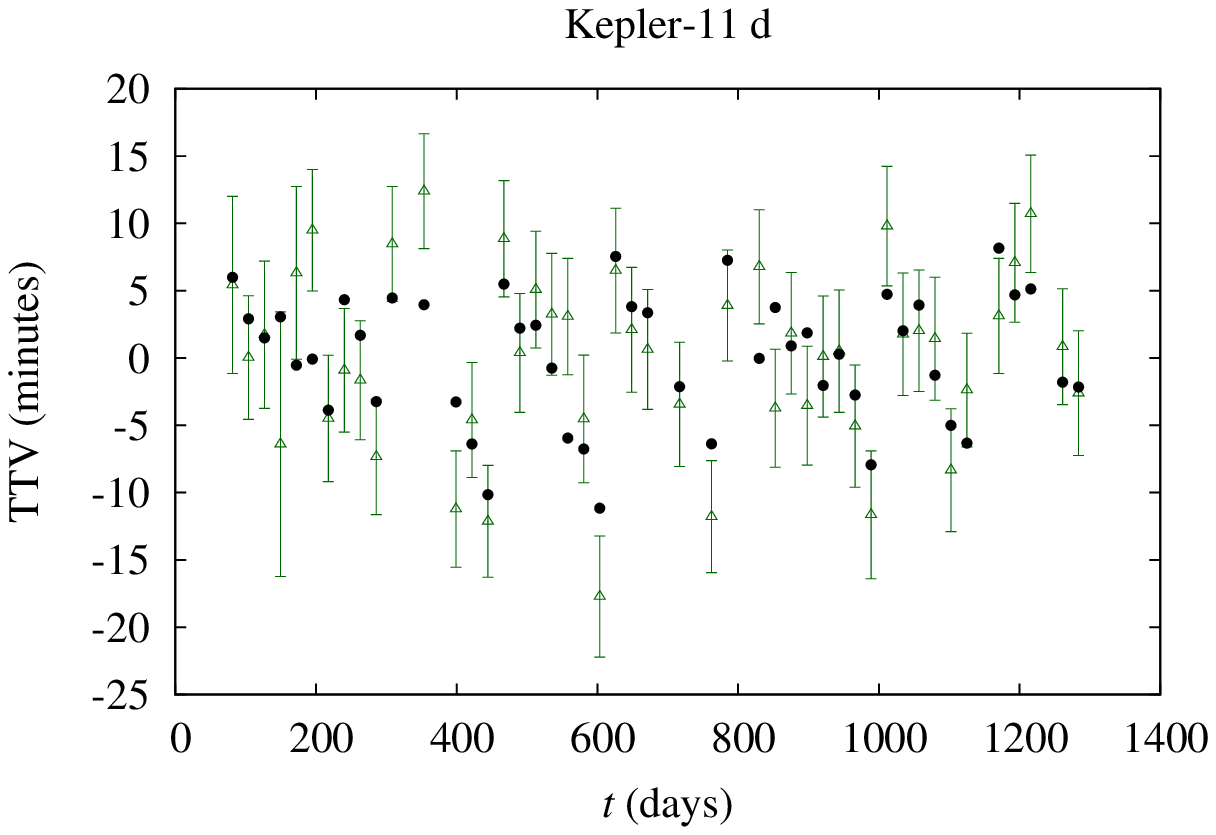}
\includegraphics [height = 2.1 in]{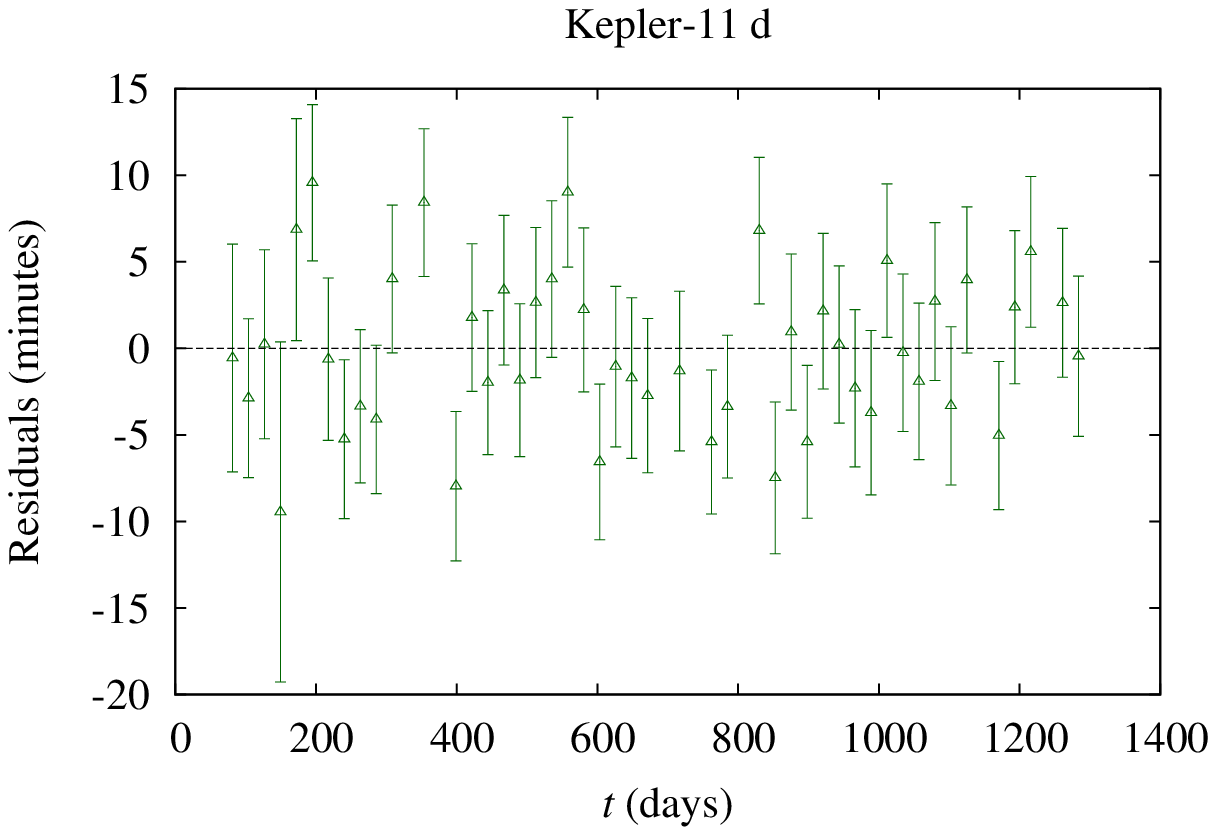}
\caption{Observed and simulated transit timing variations for Kepler-11 b, c and d, using transit time measurements from J.R. See the caption to Figure 2 for details.
}
\label{fig:JR1} 
\end{figure}

\begin{figure}
\includegraphics [height = 2.1 in]{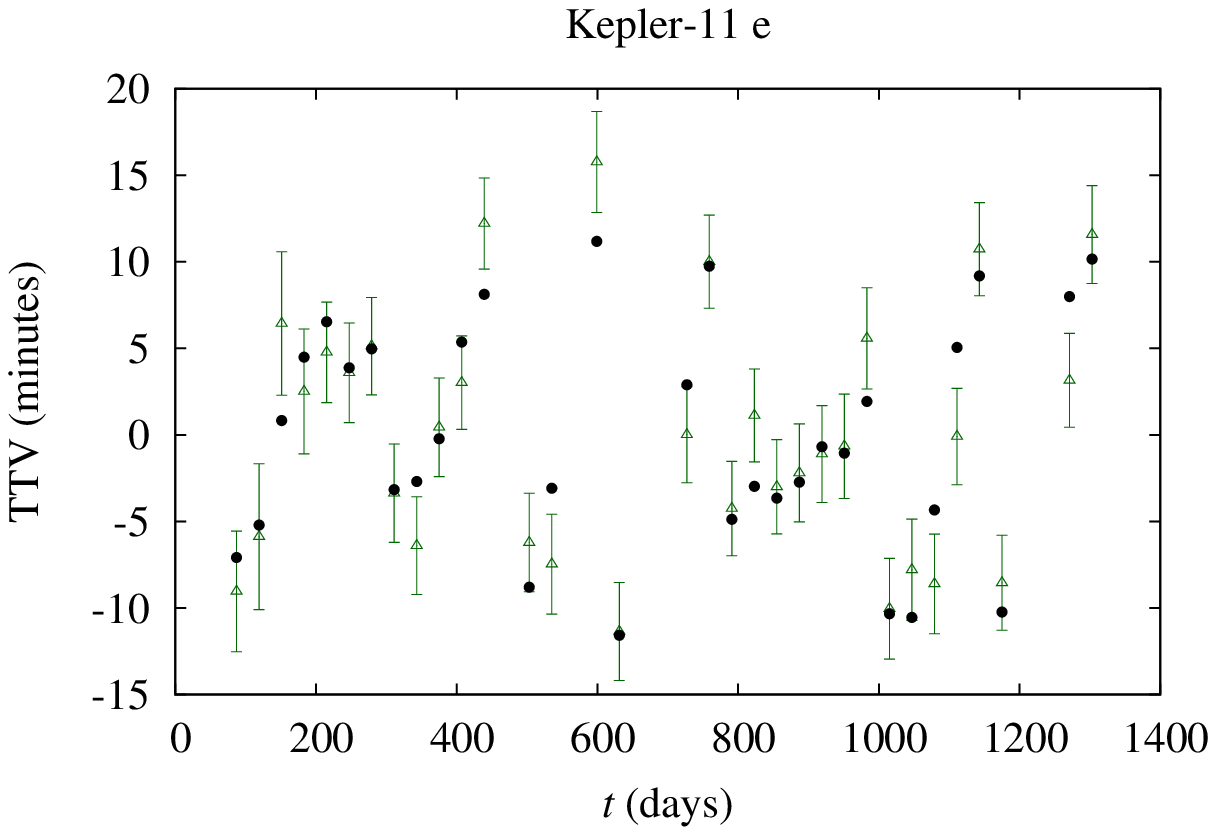}
\includegraphics [height = 2.1 in]{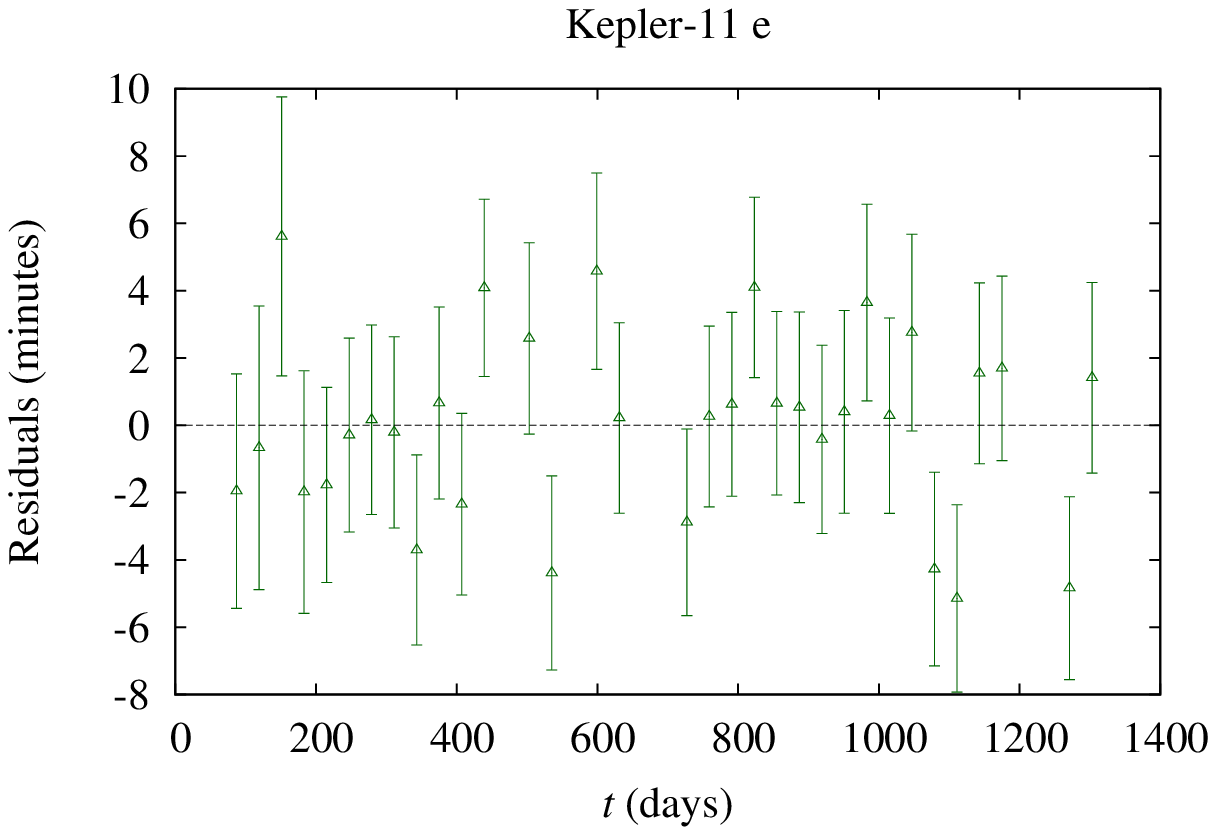}
\newline
\includegraphics [height = 2.1 in]{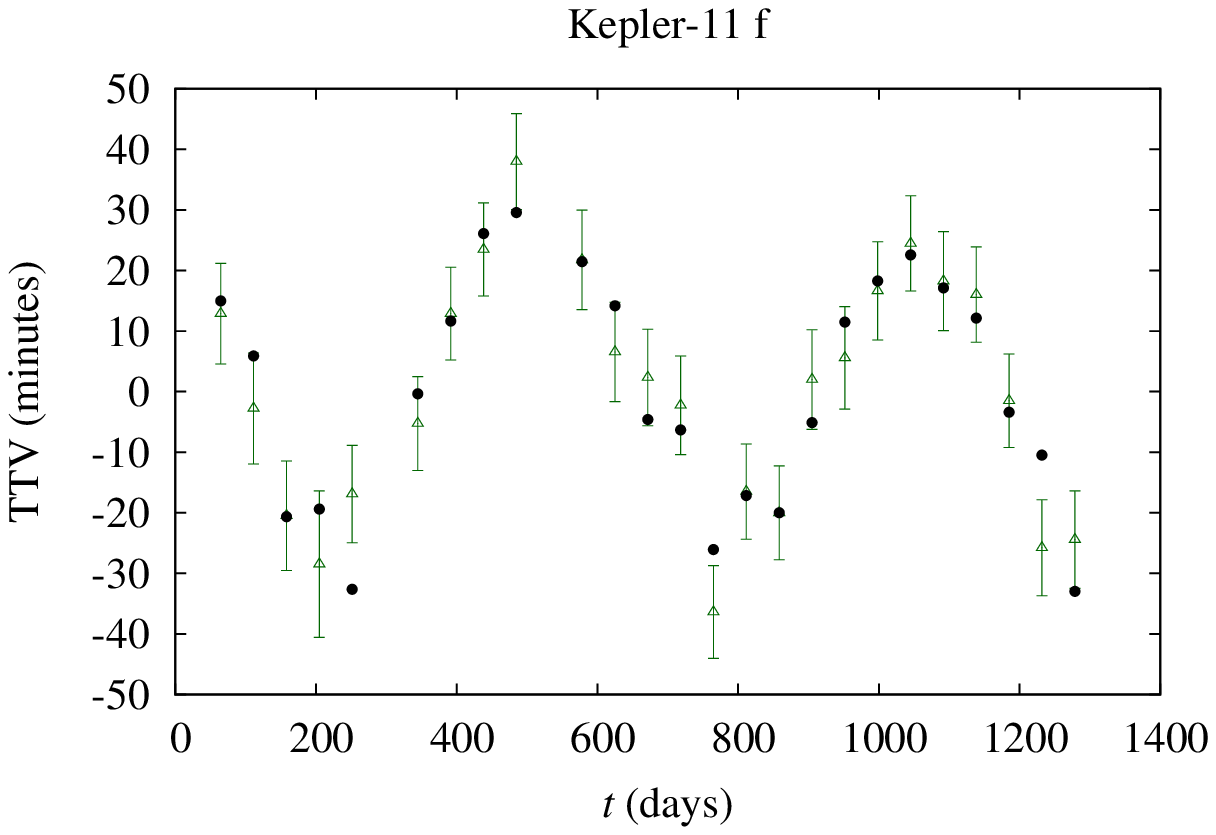}
\includegraphics [height = 2.1 in]{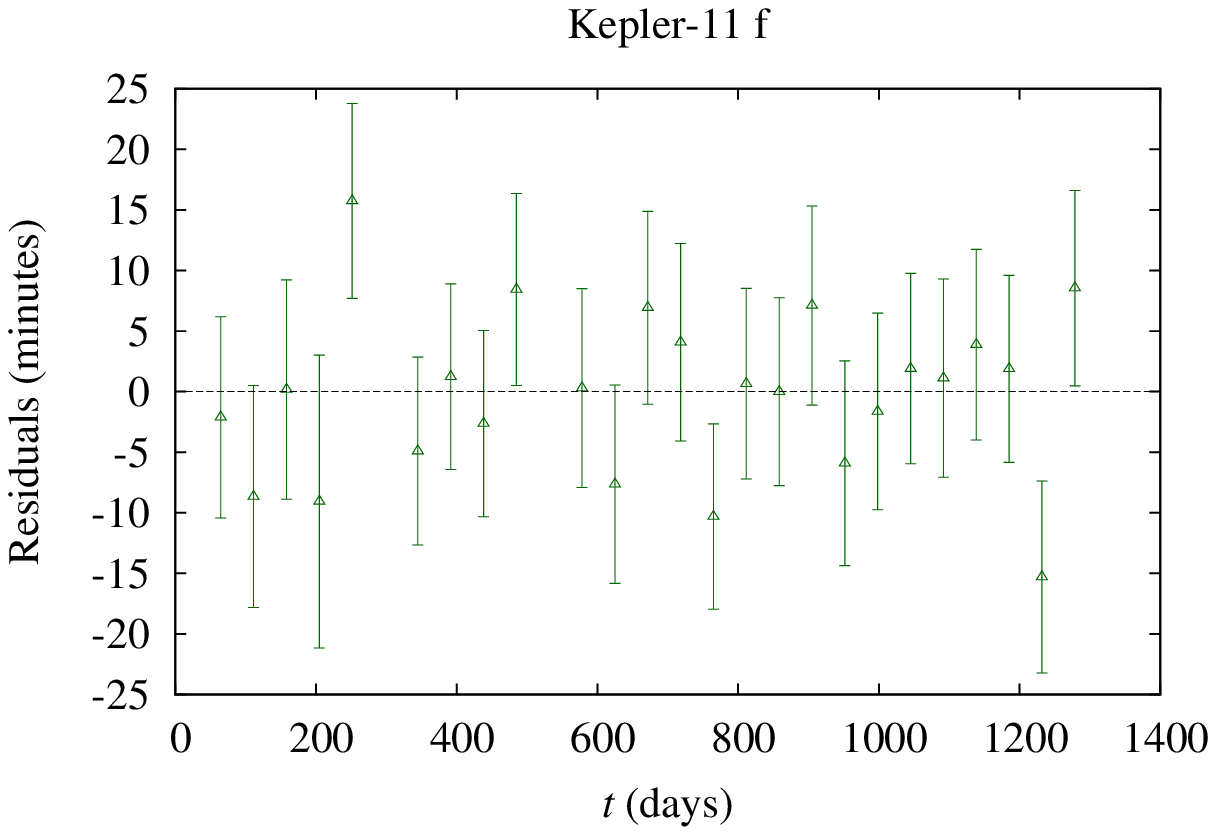}
\newline
\includegraphics [height = 2.1 in]{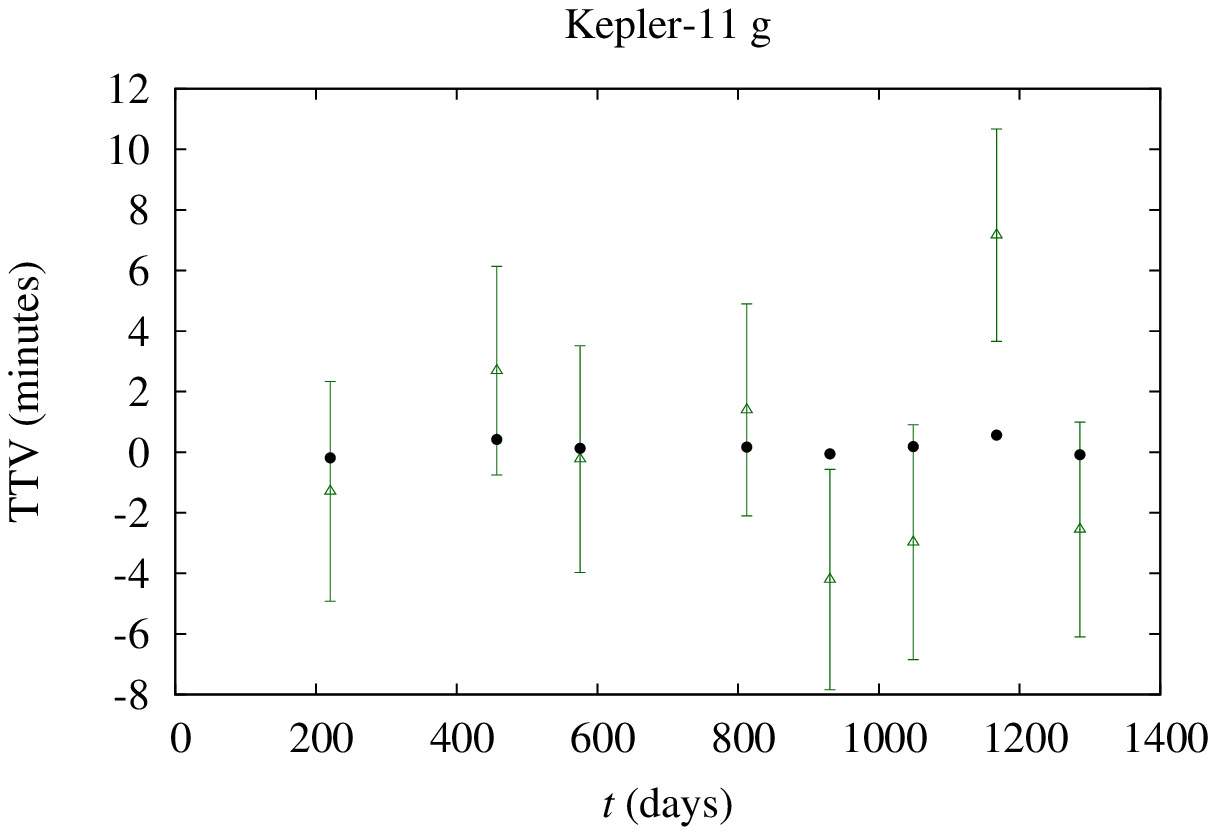}
\includegraphics [height = 2.1 in]{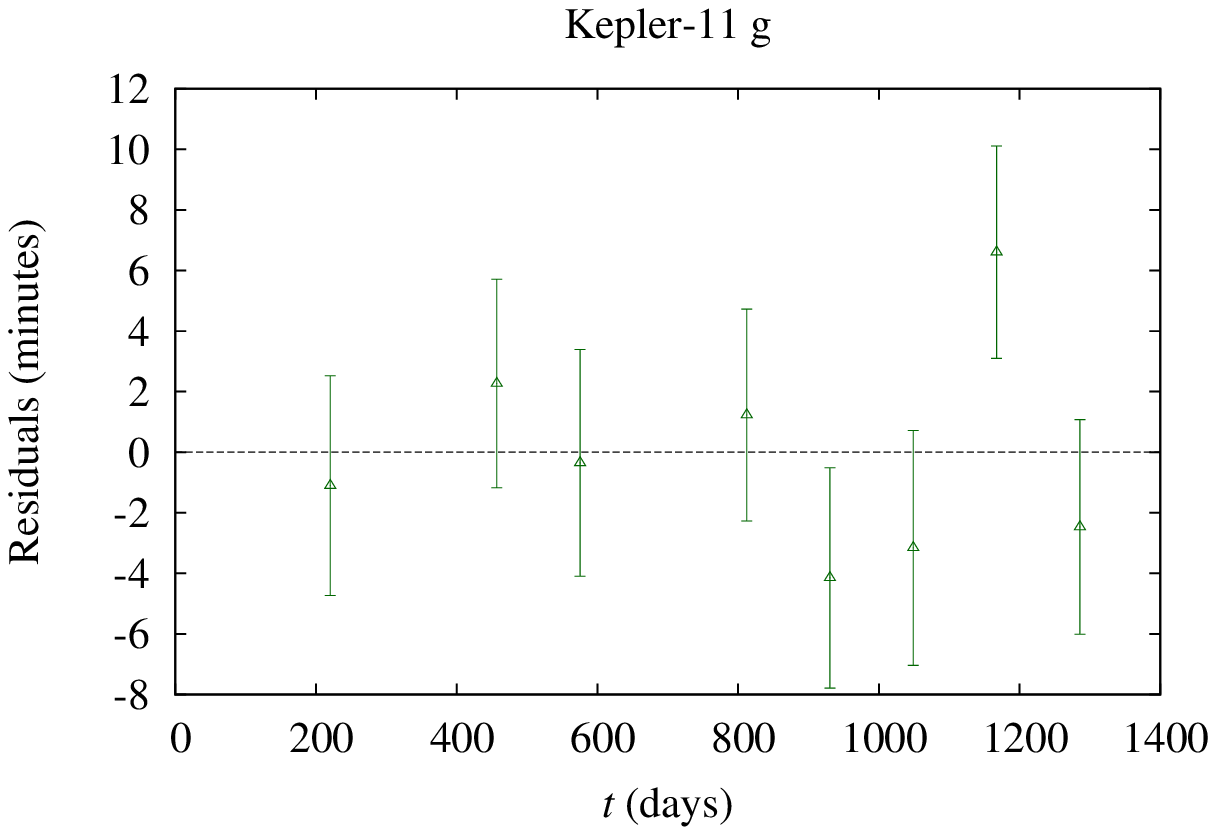}
\caption{Observed and simulated transit timing variations for Kepler-11 e, f and g, using transit time measurements from J.R.  See the caption to Figure 2 for details.}
\label{fig:JR2} 
\end{figure}

\begin{figure}
\includegraphics [height = 2.1 in]{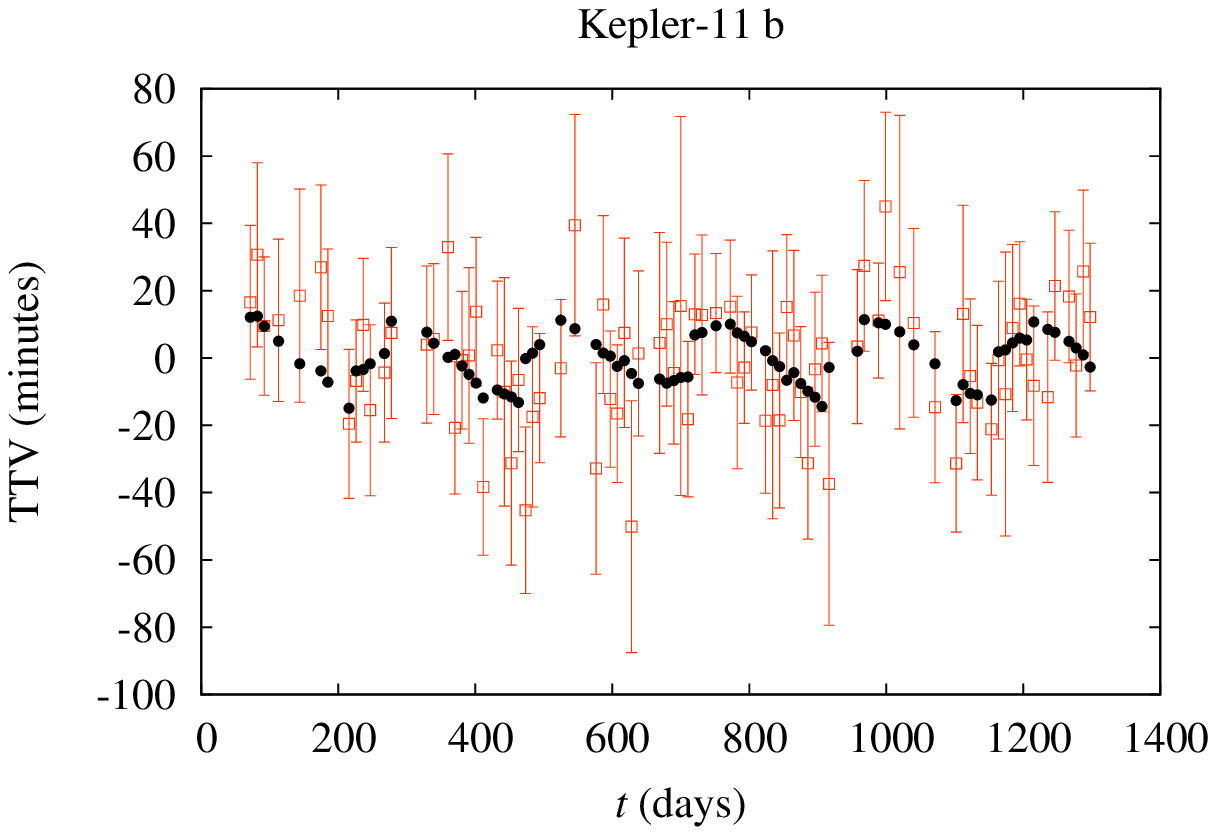}
\includegraphics [height = 2.1 in]{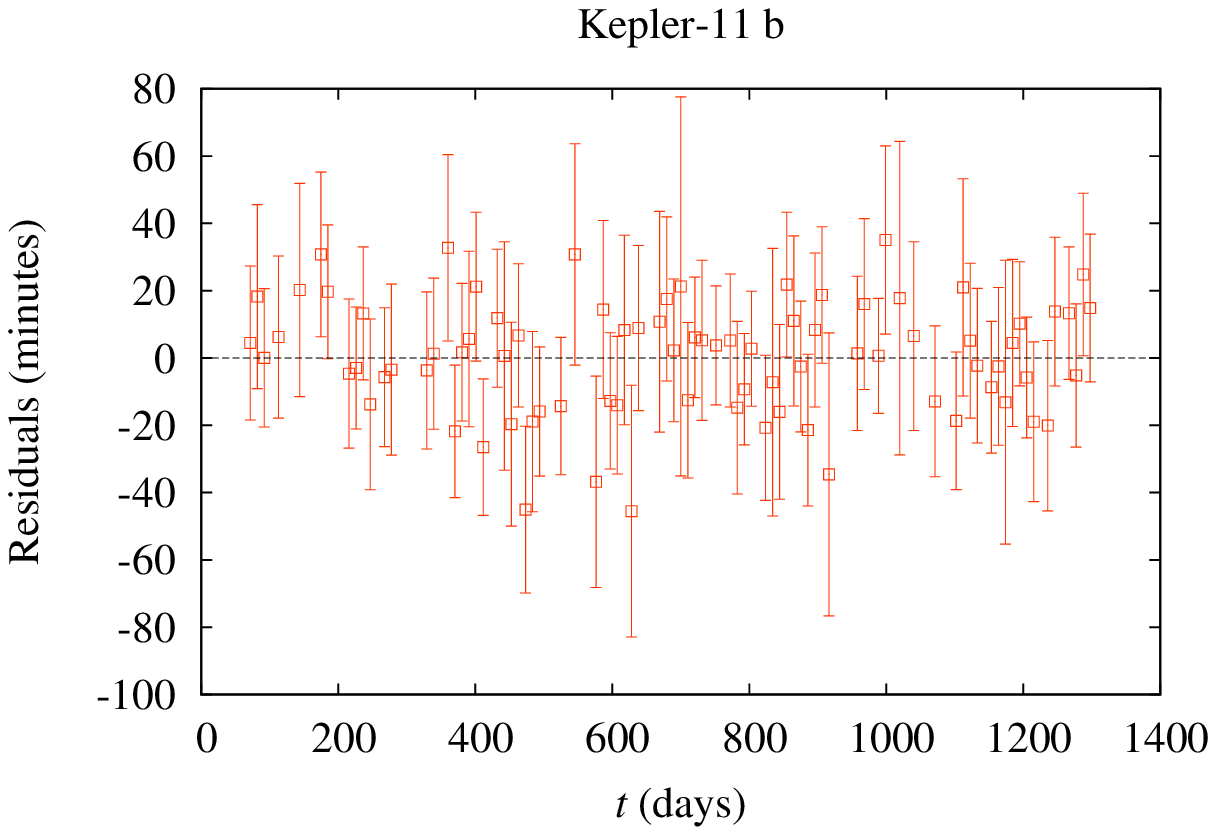}
\newline
\includegraphics [height = 2.1 in]{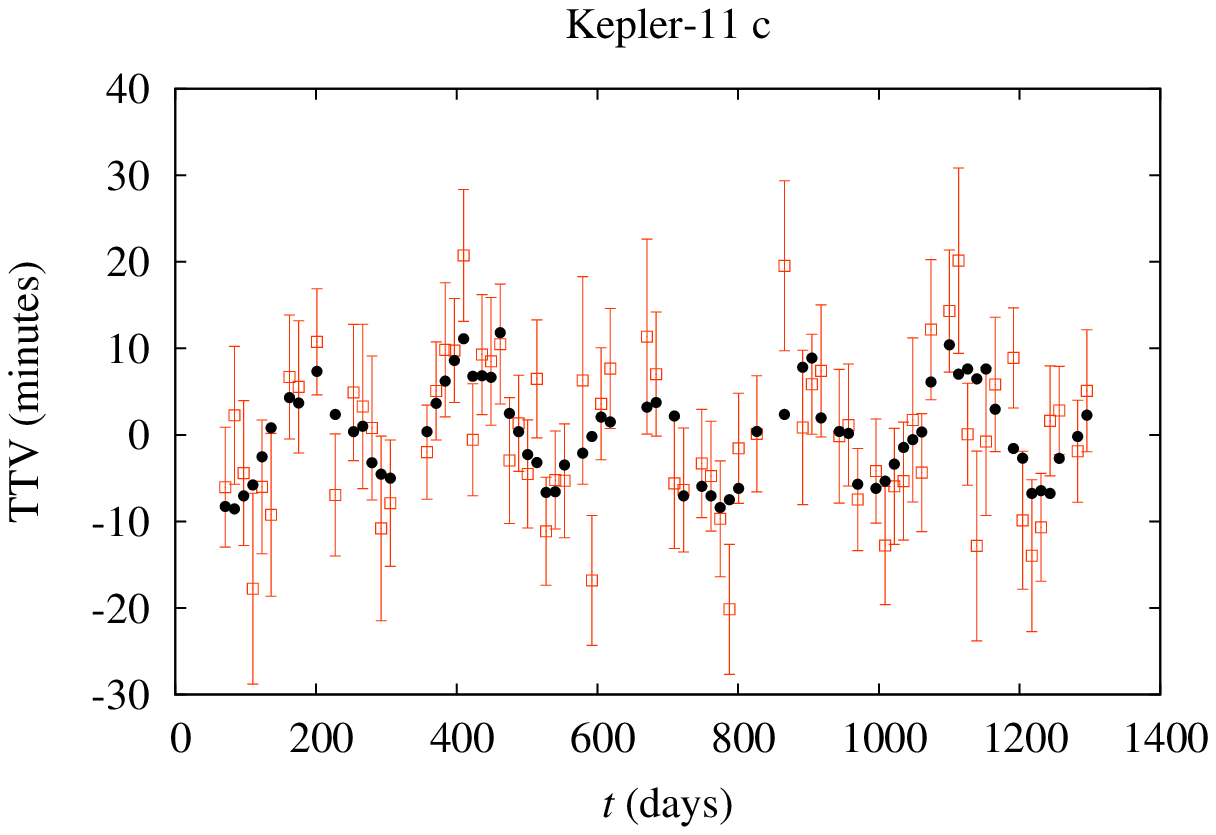}
\includegraphics [height = 2.1 in]{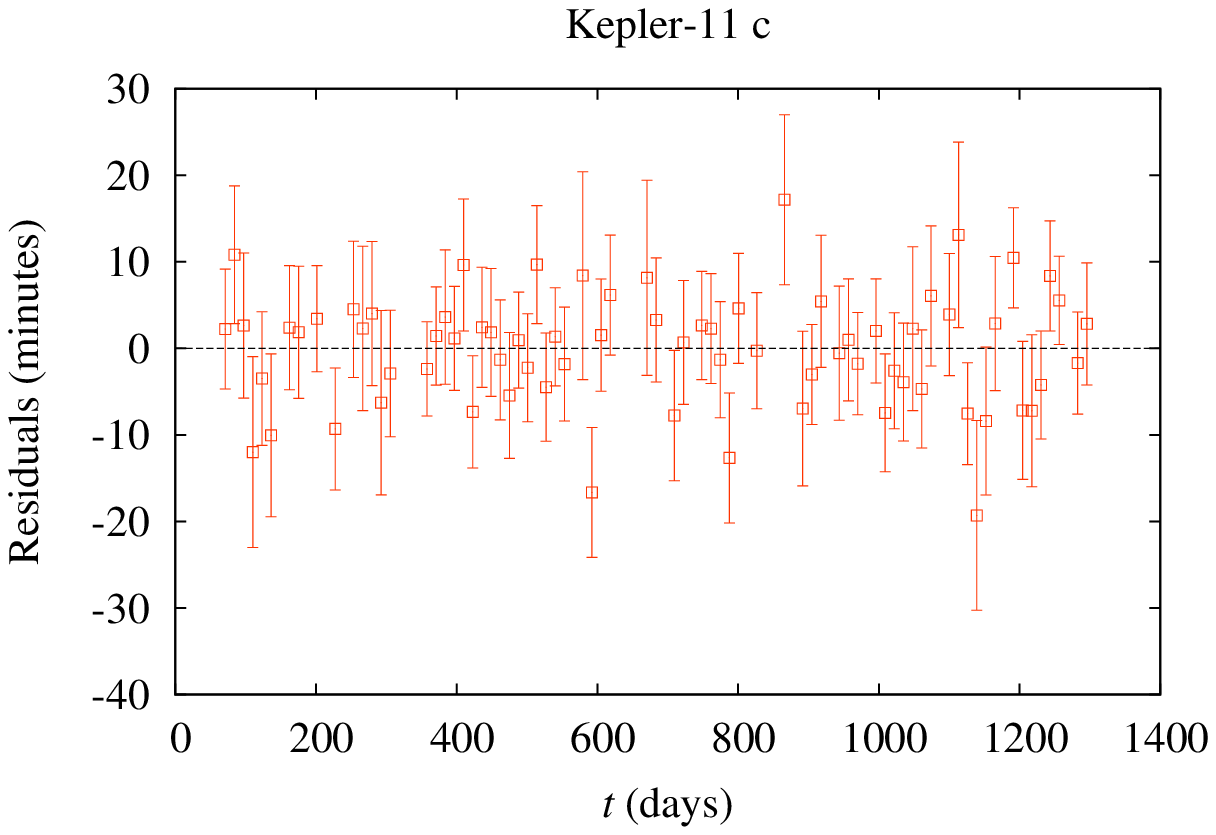}
\newline
\includegraphics [height = 2.1 in]{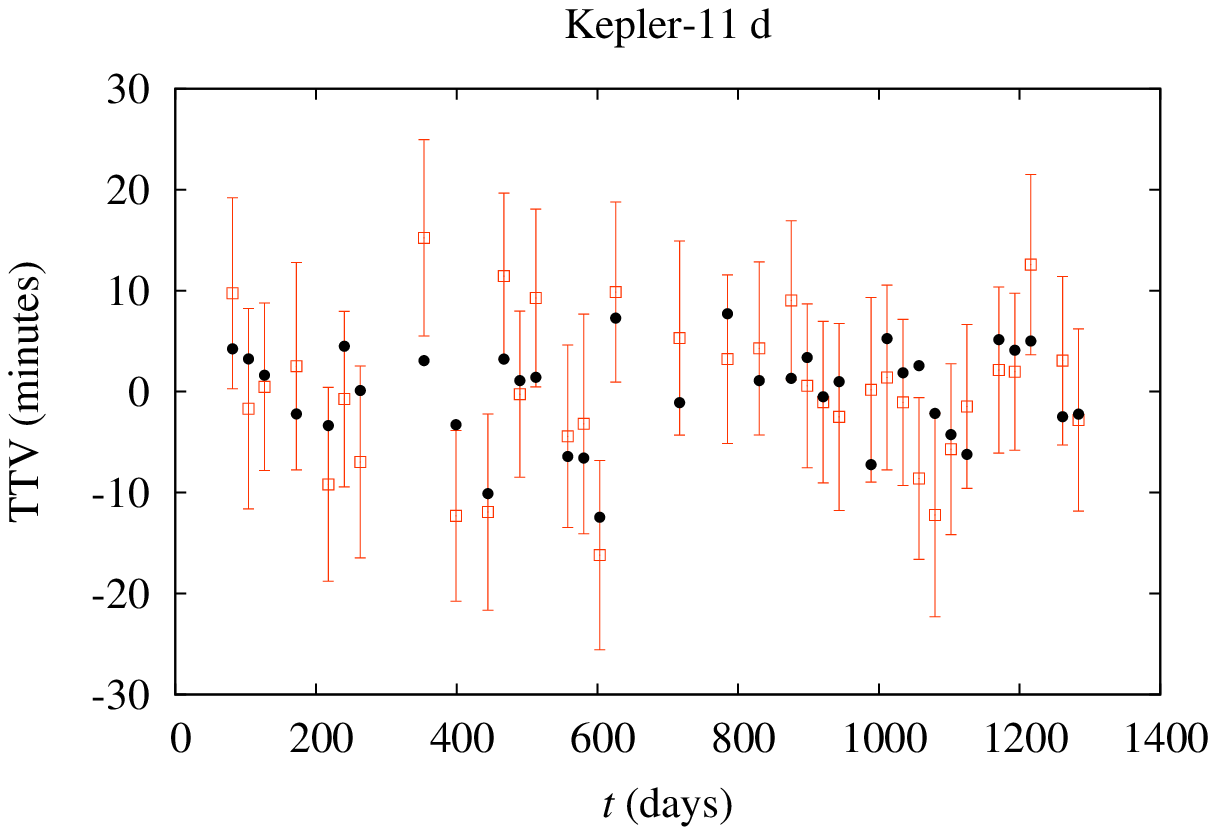}
\includegraphics [height = 2.1 in]{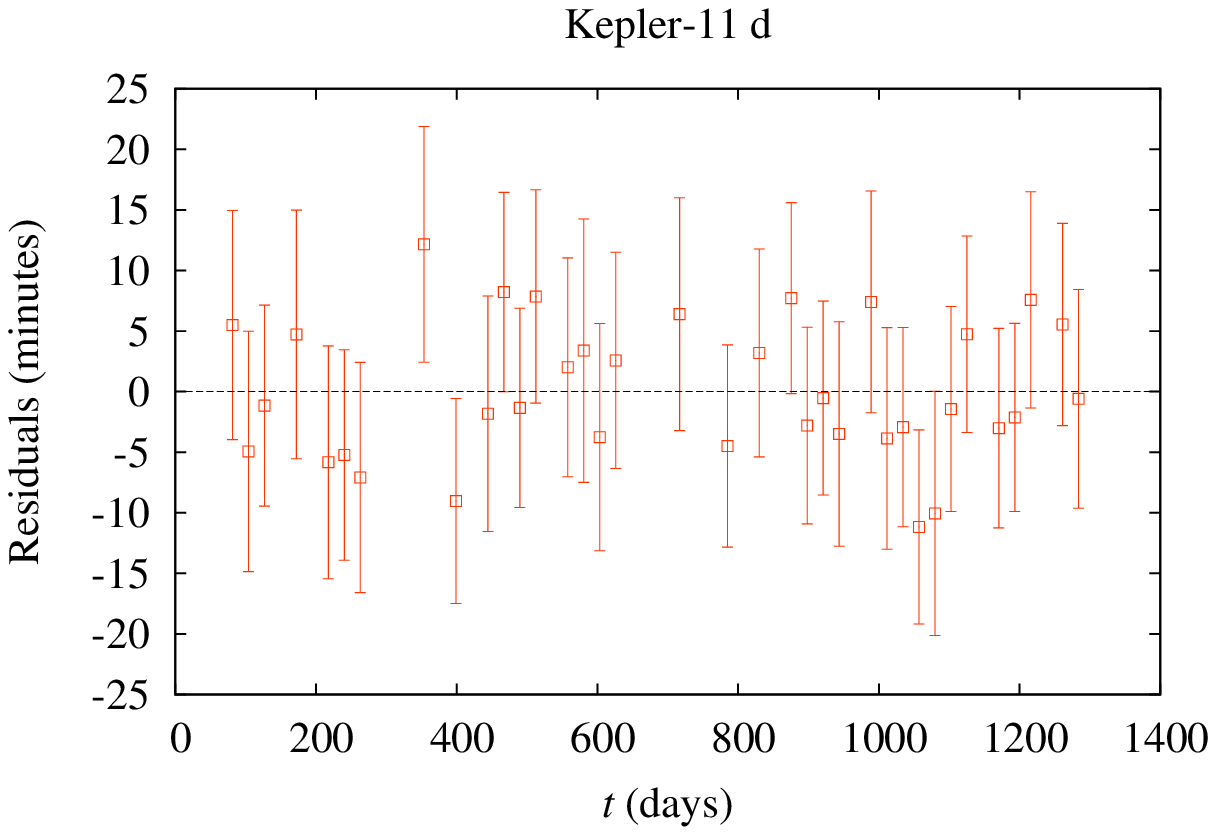}
\caption{Observed and simulated transit timing variations for Kepler-11 b, c and d, using transit time measurements from D.S. See the caption to Figure 2 for details.
}
\label{fig:DS1} 
\end{figure}

\begin{figure}
\includegraphics [height = 2.1 in]{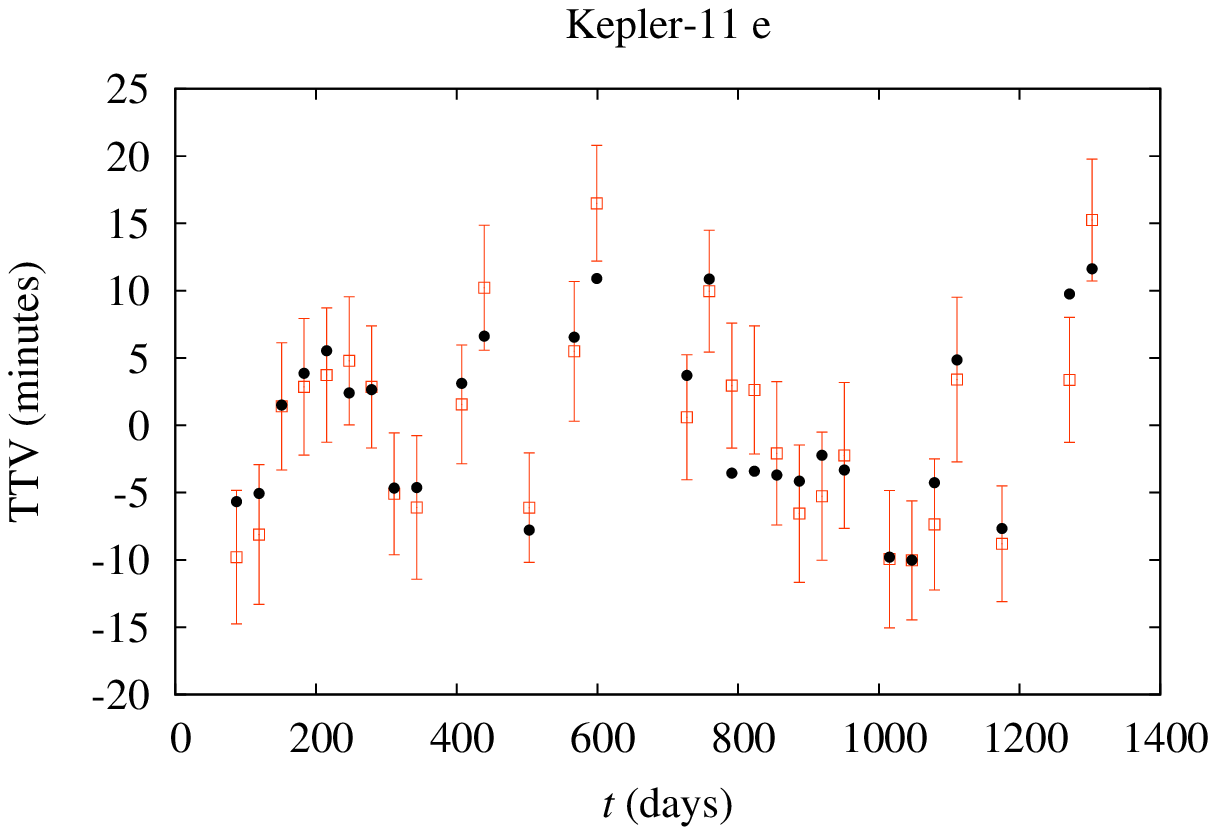}
\includegraphics [height = 2.1 in]{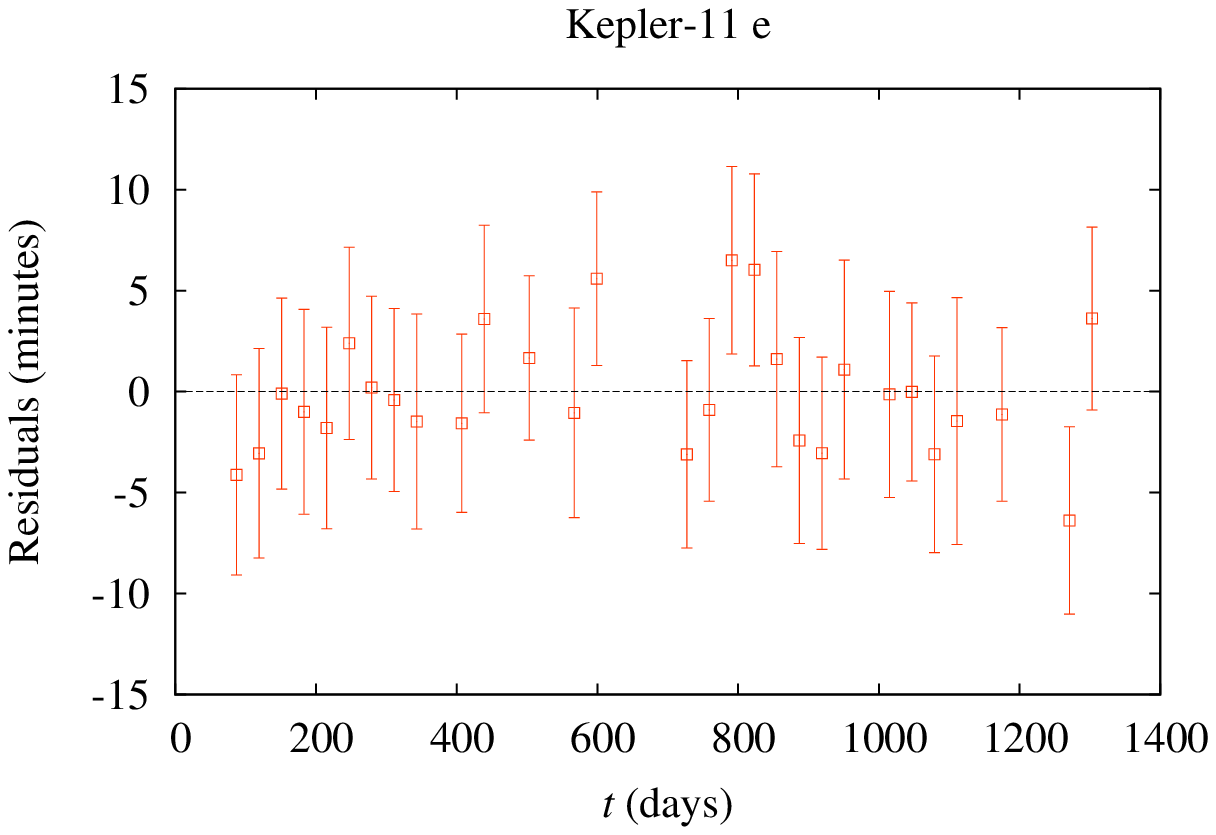}
\newline
\includegraphics [height = 2.1 in]{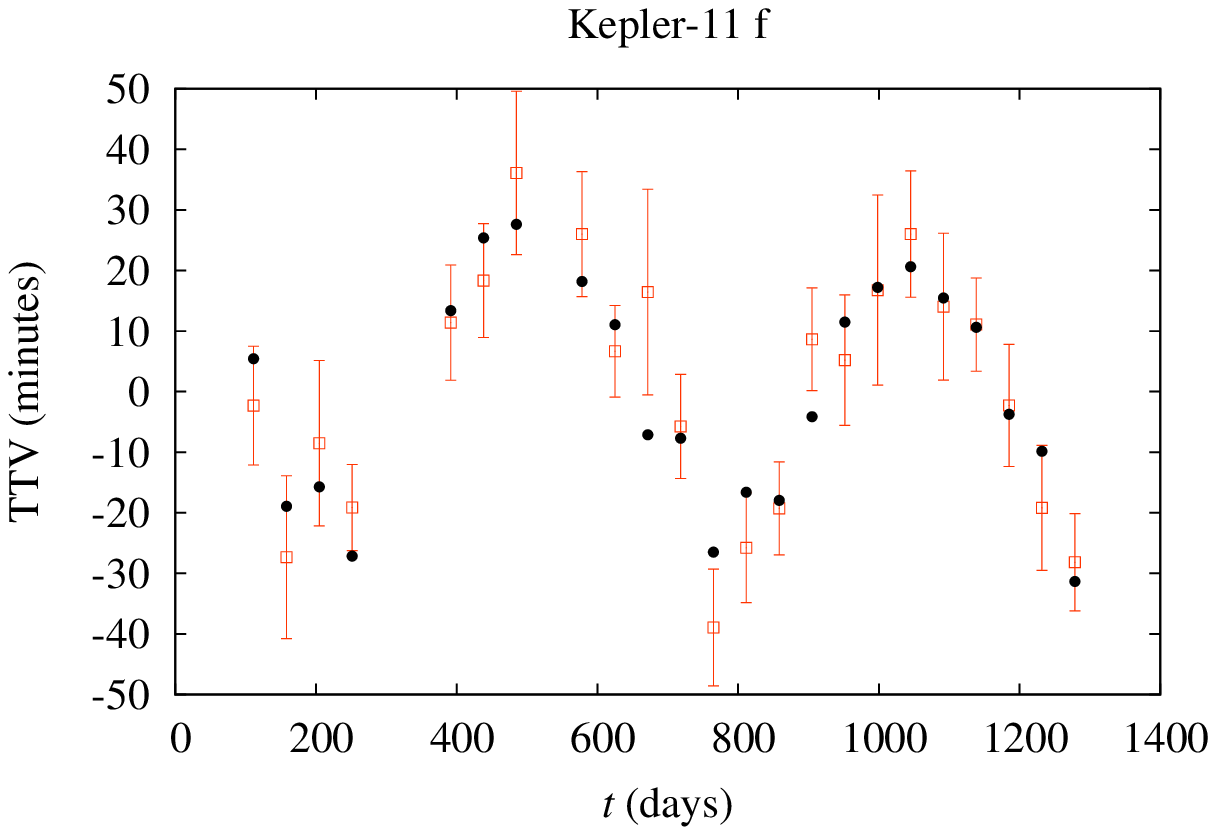}
\includegraphics [height = 2.1 in]{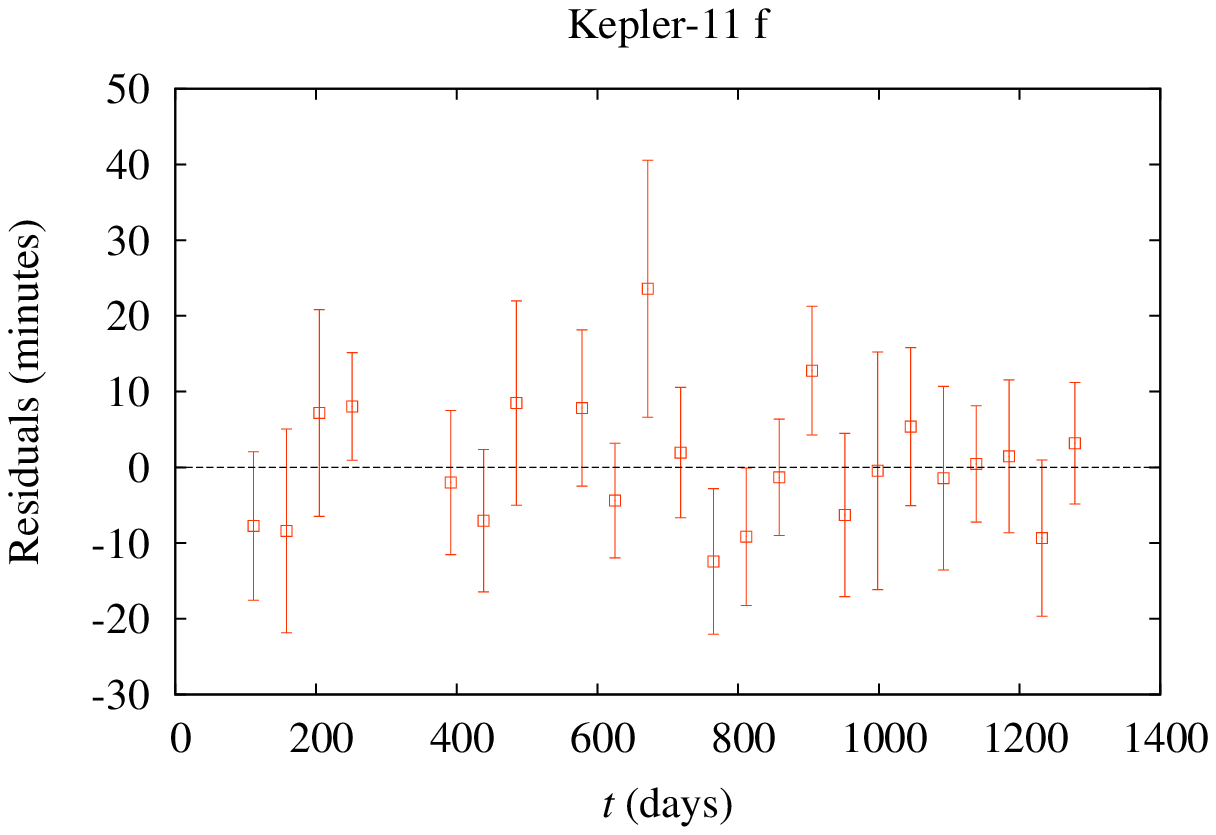}
\newline
\includegraphics [height = 2.1 in]{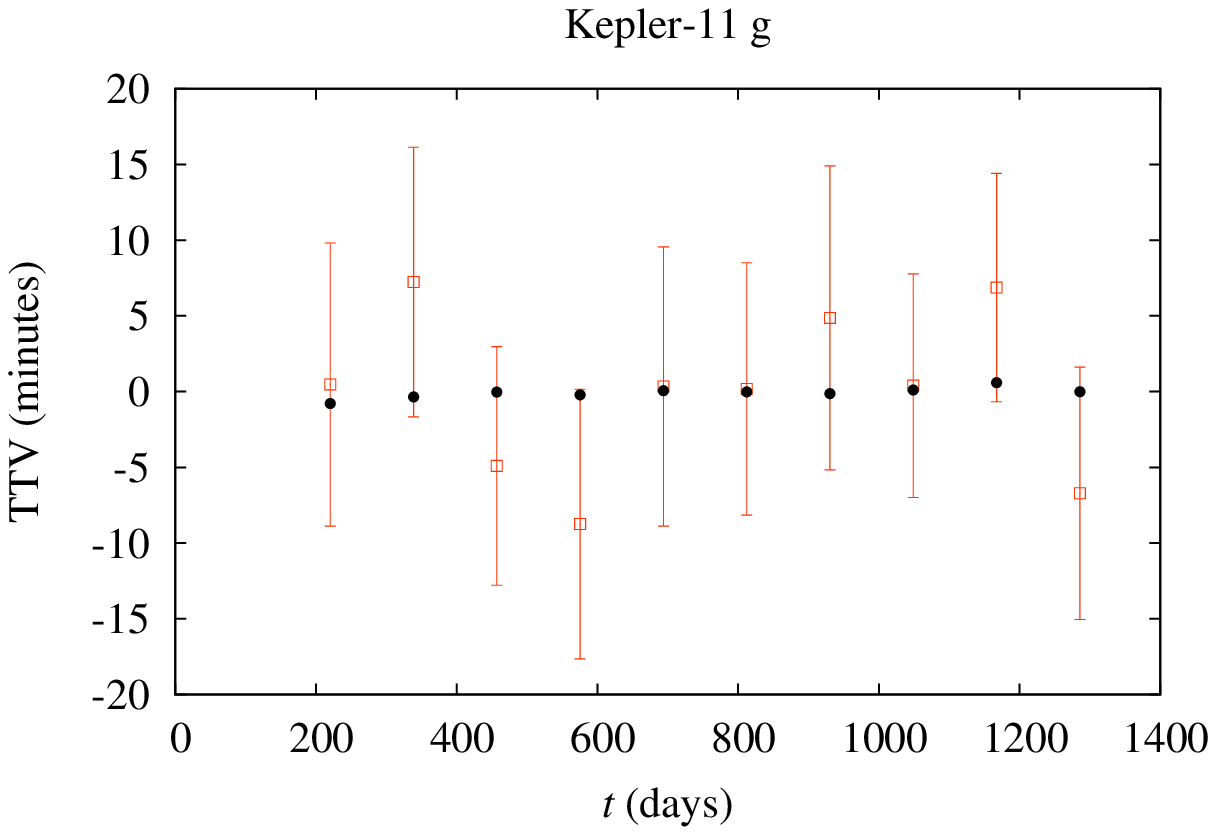}
\includegraphics [height = 2.1 in]{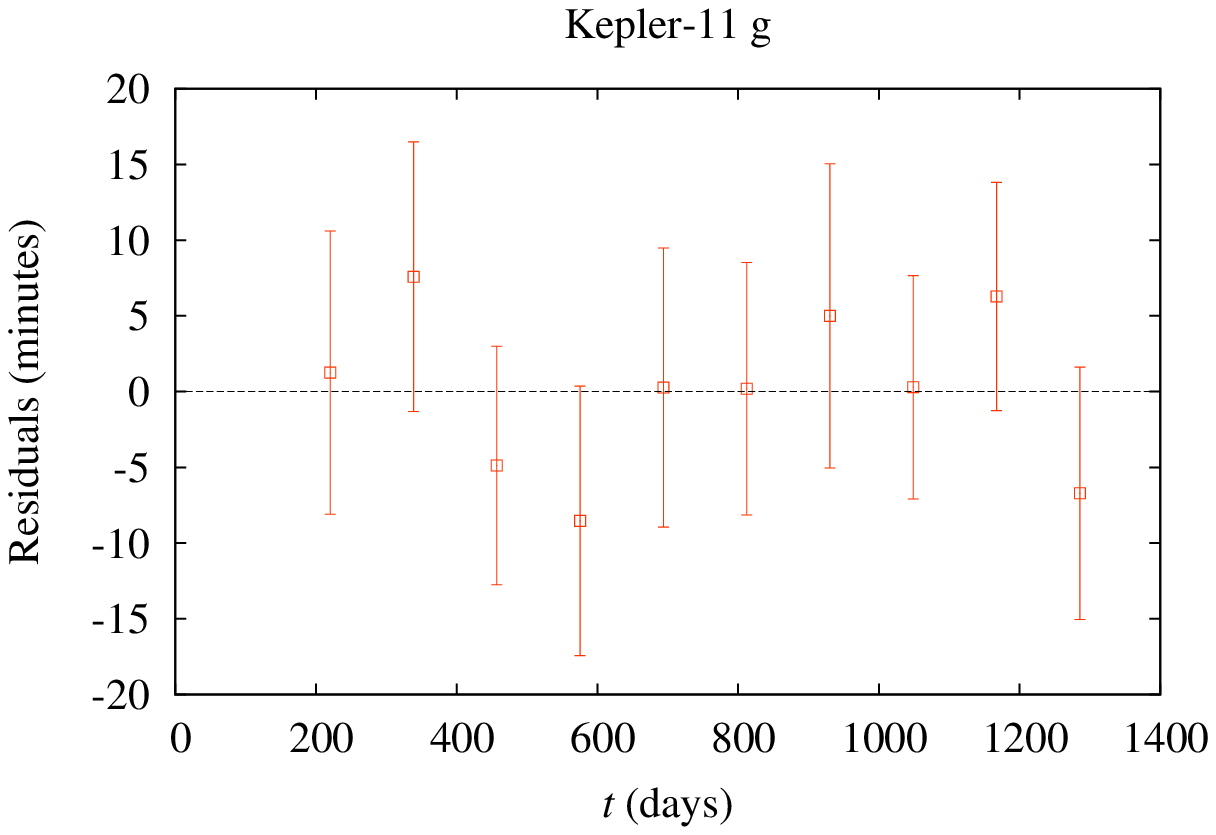}
\caption{Observed and simulated transit timing variations for Kepler-11 e, f and g, using transit time measurements from D.S.  See the caption to Figure 2 for details.}
\label{fig:DS2} 
\end{figure}

We adopted a wide variety of initial conditions for comparison, and found that our solutions were insensitive to the mass of the outer planet, Kepler-11 g. Hence for all subsequent simulations used to determine the masses and orbital parameters of the five inner planets, we keep the mass of Kepler-11 g as a fixed parameter set to $2.53 \times 10^{-5}M_\star$ (comparable to the masses of similar size planets in this system and equal to 8 M$_{\oplus}$ for the value of stellar mass estimated by \citealt{liss11a}), with its orbital eccentricity fixed at zero. We find that the masses and orbital parameters of planets Kepler-11 b--f converge to the values listed in Tables~\ref{tbl-EAbestfit} -- \ref{tbl-DSbestfit} (Appendix B), and the resulting modeled TTVs fit the data well, as displayed in Figures~\ref{fig:EA1} -- \ref{fig:DS2}.

\begin{table}
  \begin{center}
    \begin{tabular}{|cccccc|}
      \hline
      Planet  &  $P$ (days)  & $T_{0}$ (date) &   $e \cos \omega$ & $e \sin \omega$   & $M_p/M_{\star} \times 10^{-6}$ \\
      \hline
 b  & \textbf{10.3039}$^{+0.0006}_{-0.0010}$ & \textbf{689.7378}$^{+0.0026}_{-0.0047}$ &  \textbf{0.032}$^{+0.036}_{-0.032}$ &  \textbf{0.032}$^{+0.059}_{-0.029}$ & \textbf{5.84}$^{+4.25}_{-3.10}$ \\  
 c  & \textbf{13.0241}$^{+0.0013}_{-0.0008}$ & \textbf{683.3494}$^{+0.0014}_{-0.0019}$ &  \textbf{0.016}$^{+0.033}_{-0.025}$ &  \textbf{0.020}$^{+0.053}_{-0.029}$ & \textbf{9.19}$^{+9.12}_{-4.90}$ \\  
 d  & \textbf{22.6845}$^{+0.0009}_{-0.0009}$ & \textbf{694.0069}$^{+0.0022}_{-0.0014}$ &  \textbf{-0.003}$^{+0.005}_{-0.005}$ &  \textbf{0.002}$^{+0.006}_{-0.002}$ & \textbf{22.86}$^{+2.58}_{-4.83}$ \\  
 e  & \textbf{31.9996}$^{+0.0008}_{-0.0012}$ & \textbf{695.0755}$^{+0.0015}_{-0.0009}$ &  \textbf{-0.008}$^{+0.004}_{-0.003}$ &  \textbf{-0.009}$^{+0.005}_{-0.005}$ & \textbf{24.87}$^{+4.84}_{-6.68}$ \\  
 f  & \textbf{46.6888}$^{+0.0027}_{-0.0032}$ & \textbf{718.2710}$^{+0.0041}_{-0.0038}$ &  \textbf{0.011}$^{+0.009}_{-0.008}$ &  \textbf{-0.005}$^{+0.006}_{-0.007}$ & \textbf{6.32}$^{+2.63}_{-2.94}$ \\  
 g  & \textbf{118.3807}$^{+0.0010}_{-0.0006}$ & \textbf{693.8021}$^{+0.0030}_{-0.0021}$ &  (0) & (0)  & \textbf{$\lesssim 70$} \\ 
       \hline
    \end{tabular}
    \caption{Our combined fit dynamical model to the observed transit times, with the orbital periods (second column), time of first transit after JD = 2,454,900 (third column), $e\cos\omega$ (fourth column), $e\sin\omega$ (fifth column), and planetary mass in units of the stellar mass (sixth column), all as free variables for planets Kepler-11 b-f. Periods are given as viewed from the barycenter of our Solar System. Because Kepler-11 is moving towards the Solar System at 57 km/s, actual orbital periods in the rest frame of Kepler-11 are a factor of 1.00019 times as long as the values quoted (as noted by \citealt{liss11a}). The simulations used to derive these parameters adopted a circular orbit and a fixed mass of $25.3 \times 10^{-6} M_{\star}$ for Kepler-11 g. The upper limit on the mass of planet g was explored separately, as described in the text\label{tbl-dyn}.}
  \end{center}
\end{table}




Our dynamical fitting of the planetary parameters minimizes residuals by adjusting parameters to search for a best-fit, which is determined by a local minimum value of $\chi^2$.  Uncertainties are based on the assumption that the shape of the $\chi^2$ surface is well-approximated by local gradients near the minimum, i.e., is shaped like a parabola.  For multi-variate problems such as this, the dimensionality of phase space is large, and multiple minima typically exist.  Furthermore, the low S/N of some lightcurves, particularly, Kepler-11 b, makes the $\chi^2$ surface fairly rough, with many local minima.   Thus, the minimum that the code finds need not be the global minimum, i.e., the best fit to the data. And even if it does converge to the global minimum, parameters that yield other minima with $\chi^2$ only slightly larger than that of the global minimum are almost as likely to approximate well the true parameters of the system as are those of the global minimum. To qualitatively account for the increased uncertainty caused by these concerns, we combined the solutions with the three data sets by averaging their nominal values and defining error bars such that they extend over the entire range given by the union of the 1$\sigma$ confidence intervals of all three solutions; error bars are thus asymmetric. 
 Note that this gives fairly large ranges, and thus more conservative values than standard $1\sigma$ ranges - this is to compensate for shortcomings of Levenberg-Marquardt  fitting of 
such a complex multi-parameter space.

The principal results of our dynamical analysis are presented in Table~\ref{tbl-dyn}. These dynamical measurements are combined with estimates of the star's mass and radius to yield measurements of the planetary characteristics that we present in Section 5. We also performed fits to each of the three sets of TTs in which both the eccentricity and the mass of Kepler-11 g were allowed to float, as well as fits in which the mass of planet g was a free parameter but it was constrained to be on a circular orbit. In all six cases, the fits converged to values similar to those in our fits with planet g on a circular orbit at the nominal mass, albeit with large uncertainties in g's mass. When the eccentricity of planet g was allowed to float,  all six  fits were inferior (in a $\chi^2/$d.o.f. sense, where d.o.f. stands for degrees of freedom) to fits with g's parameters fixed.

\begin{figure}
\includegraphics [height = 3.1 in]{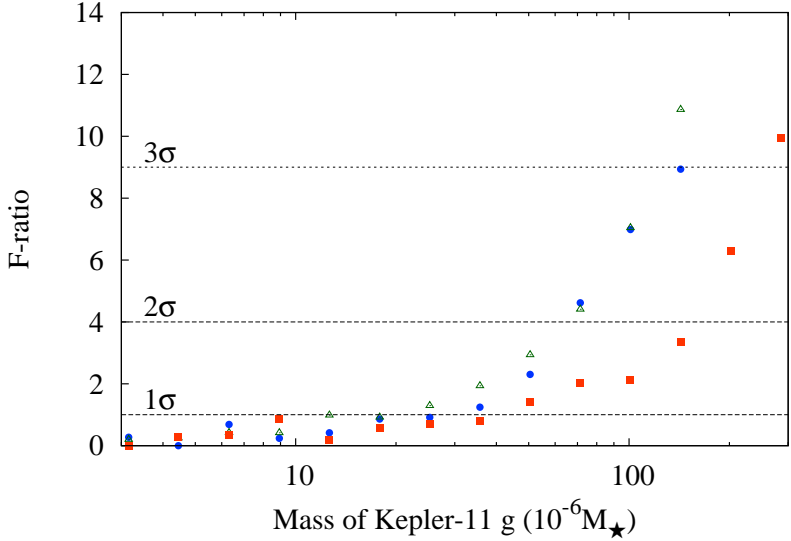}
\caption{The goodness of fit of our dynamical model to the observed TTs is shown as a function of the mass of planet Kepler-11 g. For each point, the $\chi^2$ minimum was found keeping the time of the first transit after epoch, orbital periods, eccentricities and masses as free variables for planets Kepler-11 b-f. For Kepler-11 g, the time of its first transit after epoch and its orbital period were free parameters, with its eccentricity fixed at zero, and its mass fixed in each numerical run. The vertical axis marks the F-ratio, described by Equation~(\ref{eqn:f-ratio}). Results are shown for the A.E. data with open green triangles, for the J.R. dataset in solid blue circles, and for the D.S. dataset in filled red squares. The horizontal lines mark the confidence intervals that $\chi^2$ is not elevated by chance. For the 2$\sigma$ limit, two of the datasets constrain the mass of planet g below $\sim 70 \times 10^{-6}M_{\star}$ with 95\% confidence.  (The dataset yielding weaker constraints appears to have overestimated uncertainties in measured TTs; see Table 5 and associated text.)}
\label{fig:mass-g} 
\end{figure}

To constrain the mass of Kepler-11 g, we performed a suite of simulations using the same initial conditions as our best fit to each set of transits times (see Tables~\ref{tbl-EAbestfit}, ~\ref{tbl-JRbestfit}, ~\ref{tbl-DSbestfit}).  Eccentricities for all planets except g were allowed to float in these fits, but g's eccentricity was always fixed at zero, since eccentricity and mass are inversely correlated and our goal is to determine an upper bound on Kepler-11 g's mass. For each simulation, the mass of planet g was fixed, but since we are comparing simulations with differing masses of planet g, we are effectively allowing this parameter to vary, thereby adding one degree of freedom above those in our best fit models. The F-ratio, defined as
\begin{equation}
{\rm F-ratio} = \frac{\Delta \chi^2/\Delta({\rm d.o.f.})}{\chi^2/({\rm d.o.f.})},
\label{eqn:f-ratio}
\end{equation}
describes the likelihood that a change in the minimum of $\chi^2$ could happen by chance given a change in the number of degrees of freedom, in our case, by varying the (fixed for any given run but changed from one run to another) the mass of Kepler-11 g between fits. Figure~\ref{fig:mass-g} shows the change in $\chi^2$ with variations in the mass of planet g. The $2\sigma$ limits constrain the mass of g, with a confidence of $95\%$, such that $M_{p}$(g) $\lesssim 70\times 10^{-6}~M_{\star}$ for two of the three datsets (the  error bars in third dataset, for which the mass constraint is looser, are  likely to be significantly over-estimated; see Table 5 and associated text for details).

We next consider the dynamical evolution of the Kepler-11 system using the parameters that we have derived and presented in Table 1. Our analysis treats the planets and star as point masses and neglects relativistic effects, so we do not need to know the sizes of the objects nor the mass of the star for this analysis.

One may ask whether as compact a planetary system as Kepler-11 is dynamically stable on gigayear timescales. We performed a numerical simulation of a system consisting of planets with masses and components of eccentricity equal to the nominal values in our best fit (Table~\ref{tbl-dyn}). The system remained bounded with no gross changes in orbital elements for the entire 250 Myr simulated. In contrast, an integration of a system with planetary masses and eccentricity components $1\sigma$ above the tabulated values went unstable after 1 Myr, but note that the tabulated uncertainties do not account for the anticorrelation between fitted masses and eccentricities of planets b and c, so the combination of $1\sigma$ high eccentricities and masses is highly unlikely based upon analysis of the short-term dynamics alone.  The intermediate case of a system with planetary masses and eccentricity components $0.5\sigma$ above the tabulated values went unstable after 140 Myr; however, in addition to the caveats mentioned for the  $1\sigma$ high integrations, we note that tidal damping (not included in our integrations) could well counter eccentricity growth in such a compact planetary system on $10^8$ year timescales.

We also performed precise short-term integrations of the nominal system given in Table~\ref{tbl-dyn} for 10$^7$ days using a Bulirsch-Stoer code. The eccentricities of each of the three low-mass planets, Kepler-11 b, c and f, varied from minima of $\sim 0.002 - 0.008$ to maxima between 0.04 and 0.05.  The eccentricities of Kepler-11  d and e varied from values below 0.0006 to $\sim 0.013$.  Kepler-11  g was included in these integrations, but it is weakly coupled to the other planets, and its eccentricity remained below 0.0006. We also ran an analogous integration with all planetary eccentricities 
initially set to zero.  All  eccentricities remained small, with peak 
values for the inner five planets in the range 0.0014 -- 0.0024.

\section{Properties of the Star Kepler-11}
\citet{liss11a} performed a standard SME spectroscopic analysis (\citealt{vp96,vf05}) of a high resolution (R = 60,000)  spectrum of Kepler-11 with a wavelength coverage of 360 -- 800 nm that was taken by the Keck I telescope  at  BJD = 2455521.7666 using the observing setup of the California Planet Search group \citep{mar08}. They derived an effective temperature, $T_{\rm eff} = 5680 \pm 100$ K, surface gravity, log $g = 4.3 \pm 0.2$ (cgs), metallicity, [Fe/H] = $0.0 \pm 0.1$ dex, and projected stellar equatorial rotation $v \sin i = 0.4 \pm 0.5$ km s$^{-1}$.    Combining these measurements with stellar evolutionary tracks \citep{gir00, yi01} yielded estimates of the star's mass, $M_\star = 0.95 \pm 0.10$ M$_\odot$, and radius, $R_\star = 1.1 \pm 0.1$ R$_\odot$.   

We have performed new SME analyses of the same Keck spectrum and of another spectrum of comparable quality taken with the same system at  BJD = 2455455.8028.
The combined results (weighted mean values) are:  $T_{\rm eff} = 5666 \pm 60$ K, surface gravity, log $g = 4.279 \pm 0.071$ (cgs), metallicity, [Fe/H] = $0.002 \pm 0.040$ dex, and projected stellar equatorial rotation $v \sin i = 3.86 \pm 0.85$ km/s.   These values, together with Yale-Yonsei stellar evolutionary tracks, yield estimates of the star's mass, $M_\star = 0.975 \pm 0.031$ M$_\odot$,  radius, $R_\star = 1.193 \pm 0.115$ and age = 9.7 $\pm$ 1.5 Gyr.

The TTV dynamical solution presented in Table~\ref{tbl-dyn} provides stringent constraints on the orbits of the inner five transiting planets.  We used the computed values of the planets' $e \cos\omega$ and $e \sin\omega$ shown in Table~\ref{tbl-dyn} as constraints in our transit model to provide a geometrical determination of the stellar density, \rhostar.  The transit model is similar to that described in Appendix A, but we also fit for $e\cos\omega$ and $e\sin\omega$ for each of the five inner planets.  Posterior distributions for each model parameter where estimated using a Monte-Carlo-Markov-Chain (MCMC) algorithm similar to the one that is described in \citet{for05}, but augmented with a parameter buffer to allow jumps that account for correlated variables as described in Rowe 
et al.~(2013, in preparation).  We produced 4 Markov-Chains, each with a length of 2,500,000.  We ignored the first 40\% of each chain as burn in and combined the remainder into one chain of length 6,000,000.  We adopted the median value for each model parameter, which we list in Table~\ref{tbl-transit}.

Since the dynamical model provides a good solution for the orbits of the planets from modeling of the TTVs, we reran the transit model and use the constraints on \ecosw\ and \esinw\ to estimate \rhostar.
This translates into the tight constraint: \rhostar\ = 1.122$^{+0.049}_{-0.060}$. We combined this estimate of \rhostar\ with the new (weighted mean) SME spectroscopic parameters to determine the stellar mass and radius by fitting \teff, \logg\ and \feh\ to $M_{\star}$, age and heavy element mass fraction, $Z$, as provided by the Yale-Yonsei evolution models.  We used our MCMC algorithm to determine posterior distributions of the stellar parameters and adopted the median value for each parameter as listed in Table \ref{tbl-star}. Note that the star is slightly evolved,  more than halfway through its lifetime on the main sequence.

\begin{table}[h!]
  \begin{center}
    \begin{tabular}{|ccccccc|}
   \hline
Planet & $R_p/R_{\star}$                       & duration (h)         &  depth (ppm)               & $b$                      & $i$ ($^{\circ}$)          & $a/R_{\star}$ \\ 
\hline
   b   &   0.01563$^{+0.00018}_{-0.00023}$ &  4.116$^{+0.053}_{-0.078}$  &  301.3$^{+7.3}_{-7.9}$   &  0.116$^{+0.053}_{-0.116}$ &  89.64$^{+0.36}_{-0.18}$    & 18.55$^{+0.31}_{-0.23}$  \\
   c   &   0.02496$^{+0.00015}_{-0.00019}$ &  4.544$^{+0.033}_{-0.046}$  &  750.8$^{+6.8}_{-10}$   &  0.156$^{+0.059}_{-0.156}$ &  89.59$^{+0.41}_{-0.16}$    & 21.69$^{+0.37}_{-0.27}$  \\
   d   &   0.02714$^{+0.00018}_{-0.00019}$ &  5.586$^{+0.045}_{-0.079}$  &  885.0$^{+11}_{-11}$ &  0.181$^{+0.074}_{-0.084}$ &  89.67$^{+0.13}_{-0.16}$   & 31.39$^{+0.53}_{-0.39}$  \\
   e   &   0.03643$^{+0.00021}_{-0.00028}$ &  4.165$^{+0.019}_{-0.040}$  &  1333$^{+14}_{-14}$      &  0.763$^{+0.008}_{-0.008}$ &   88.89$^{+0.02}_{-0.02}$   & 39.48$^{+0.67}_{-0.49}$   \\
   f   &   0.02169$^{+0.00026}_{-0.00026}$ &  6.431$^{+0.082}_{-0.089}$  &  548$^{+12}_{-12}$       &  0.463$^{+0.030}_{-0.032}$ &  89.47$^{+0.04}_{-0.04}$ &  50.79$^{+0.86}_{-0.63}$  \\
   g   &   0.02899$^{+0.00022}_{-0.00032}$ &  9.469$^{+0.086}_{-0.122}$  &  1006$^{+15}_{-19}$      &  0.217$^{+0.092}_{-0.087}$ &  89.87$^{+0.05}_{-0.06}$ &  94.4$^{+1.6}_{-1.2}$  \\
       \hline
        \hline
    \end{tabular}
    \caption{Transit constraints on the planets of Kepler-11, following dynamical models; $b$ signifies impact parameter, $i$ inclination of the orbit to to the plane of the sky and $a$ the orbital semimajor axis.}\label{tbl-transit}
  \end{center}
\end{table}

\begin{table}[h!]
  \begin{center}
    \begin{tabular}{||l|l||}
   \hline
\hline
$ M_{\star} (M_{\odot})$  &  0.961$^{+0.025}_{-0.025}$    \\
$ R_{\star} (R_{\odot})$  &  1.065$^{+0.017}_{-0.022}$    \\
$ L_{\star} (L_{\odot})$  &  1.045$^{+0.061}_{-0.078}$    \\
$T_{\rm eff}$ (K)        &  5663$^{+55}_{-66}$              \\
\logg\ (cm s$^{-2}$)    &  4.366$^{+0.014}_{-0.016}$  \\
$Z$                     &   0.0182$^{+0.0015}_{-0.0017}$ \\
$\rho_{\star}$ (g cm$^{-3}$) & 1.122$^{+0.049}_{-0.060}$  \\
Age (Gyr)  &   8.5$^{+1.1}_{-1.4}$ \\ 
       \hline
        \hline
    \end{tabular}
    \caption{The characteristics of the star Kepler-11, with 1$\sigma$ uncertainties.}\label{tbl-star}
  \end{center}
\end{table}

We also conducted a search for spectral evidence of a companion star. We began by fitting the observed spectrum of Kepler-11 obtained on BJD = 2455521.7666  (UT = 21 Nov 2010) with the closest-matching  (in a $\chi^2$ sense) member of our library of 800 stellar spectra.  The stars in our library have $T_{\rm eff} = 3500 - 7500$ K and $\log g  = 2.0 - 5.0$, which spans the FGK and early M type main sequence and subgiant stars. All library stars have accurate parallax measurements, allowing for good estimates of stellar mass and radius for each. The Kepler-11 spectrum is placed on a common wavelength scale and normalized in intensity. The $\chi^2$ value is then calculated as the sum of the squares of the differences between the Kepler-11 
 spectrum and each library spectrum. The final stellar properties are determined by the weighted mean of the ten library spectra with the lowest $\chi^2$ values. We adopt errors in each parameter by comparing results for standard stars.  The closest-matching spectrum is modified superficially by removing the Doppler shift relative to the observed spectrum, applying needed artificial rotational broadening, setting the continuum normalization, and diluting the line strengths (due to a possible secondary star), thereby achieving a best-fitting spectrum that can be subtracted from the observed spectrum to yield residuals.

We search for secondary stars by taking the residuals to that first spectral fit and performing the same $\chi^2$ search for a ``second'' spectrum that best fits those residuals; details will be presented in Kolbl et al.~(in preparation). Our approach assumes that spectra are single, until proven double, rather than immediately doing a self-consistent two-spectrum fit. This stems from an Occam's razor perspective; the notion is that if the target's spectrum is adequately fit by a single library spectrum, without need to invoke a second spectrum, then the target's spectrum can only be deemed single.  A minimum in $\chi^2$ as a function of Doppler shift for the fit of any library spectrum (actually a representative subset of them) to the residuals serves to indicate the presence of a second spectrum. We adopt a detection threshold that is approximately a $3 \sigma$ detection of the secondary star.

We find no stellar companion to Kepler-11 within $0.4\arcsec$ of the primary star, corresponding to half of the slit-width ($0.87\arcsec$) of the Keck-HIRES spectrometer. The detection threshold for any companion star depends on the RV separation between the primary star and
the putative secondary star. For all RV separations greater than 20 km s$^{-1}$, we would detect (at $3\sigma$) any
companions that are 2\% as bright (in the optical) as the primary star. For RV separations of 10 km s$^{-1}$, the detection threshold rises to 3\% as bright as the primary star, and for RV separations smaller than 10 km s$^{-1}$, the detection threshold rises rapidly to unity for FGK stars, but remains at 3\% for M dwarfs due to their very different spectra. The poor detectability of FGK-type companion stars having little Doppler offset is caused by overlap of the absorption lines.  
 
Speckle images for Kepler-11 show no nearby star. Neighbors located in an annulus from 0.05 to $0.7\arcsec$ from Kepler-11 would have been detected if their brightness were within 3 magnitudes in either V or I band, and those between  0.7 and $1.9\arcsec$ distant would have been seen down to a magnitude difference of 4 in either band.

\section{Properties of the Planets Orbiting Kepler-11}

\begin{table}[h!]
  \begin{center}
    \begin{tabular}{|ccccccc|}
      \hline
   \hline
      Planet  &  Mass ($M_{\oplus}$)     & Radius ($R_{\oplus}$)     &   Density (g cm$^{-3}$)       &  $a$  (AU)                            &   $e$    &     Flux ($F_{\odot, 1AU}$) \\
\hline
 b  & \textbf{1.9}$^{+1.4}_{-1.0}$     & \textbf{1.80}$^{+0.03}_{-0.05} $ & \textbf{1.72}$^{+1.25}_{-0.91}$  & \textbf{0.091}$^{+0.001}_{-0.001} $ &   \textbf{0.045}$^{+0.068}_{-0.042}$ &  \textbf{125.1} \\  
 c  & \textbf{2.9}$^{+2.9}_{-1.6}$     & \textbf{2.87}$^{+0.05}_{-0.06}$ & \textbf{0.66}$^{+0.66}_{-0.35}$   & \textbf{0.107}$^{+0.001}_{-0.001} $ &    \textbf{0.026}$^{+0.063}_{-0.013}$ &   \textbf{91.6} \\  
 d  & \textbf{7.3}$^{+0.8}_{-1.5}$     & \textbf{3.12}$^{+0.06}_{-0.07} $ & \textbf{1.28}$^{+0.14}_{-0.27}$  & \textbf{0.155}$^{+0.001}_{-0.001} $ &    \textbf{0.004}$^{+0.007}_{-0.002}$ &   \textbf{43.7} \\  
 e  & \textbf{8.0}$^{+1.5}_{-2.1}$     & \textbf{4.19}$^{+0.07}_{-0.09} $ & \textbf{0.58}$^{+0.11}_{-0.16}$  & \textbf{0.195}$^{+0.002}_{-0.002} $ &    \textbf{0.012}$^{+0.006}_{-0.006}$ &   \textbf{27.6} \\  
 f  & \textbf{2.0}$^{+0.8}_{-0.9}$     & \textbf{2.49}$^{+0.04}_{-0.07} $ & \textbf{0.69}$^{+0.29}_{-0.32}$  & \textbf{0.250}$^{+0.002}_{-0.002} $ &    \textbf{0.013}$^{+0.011}_{-0.009}$ &   \textbf{16.7} \\  
 g  & $<$ 25              & \textbf{3.33}$^{+0.06}_{-0.08} $ & $<$4            & \textbf{0.466}$^{+0.004}_{-0.004} $ &     < 0.15 &   \textbf{4.8} \\  
       \hline
        \hline
    \end{tabular}\label{tbl-planet}
    \caption{The planets of Kepler-11. The mass and eccentricity of Kepler-11 g are 2$\sigma$ upper bounds. All other uncertainties are 1$\sigma$ confidence intervals.}
  \end{center}
\end{table}

\begin{figure*}[placement h] 
  \begin{center}
    \includegraphics[width=7.0in,height=5.0in]{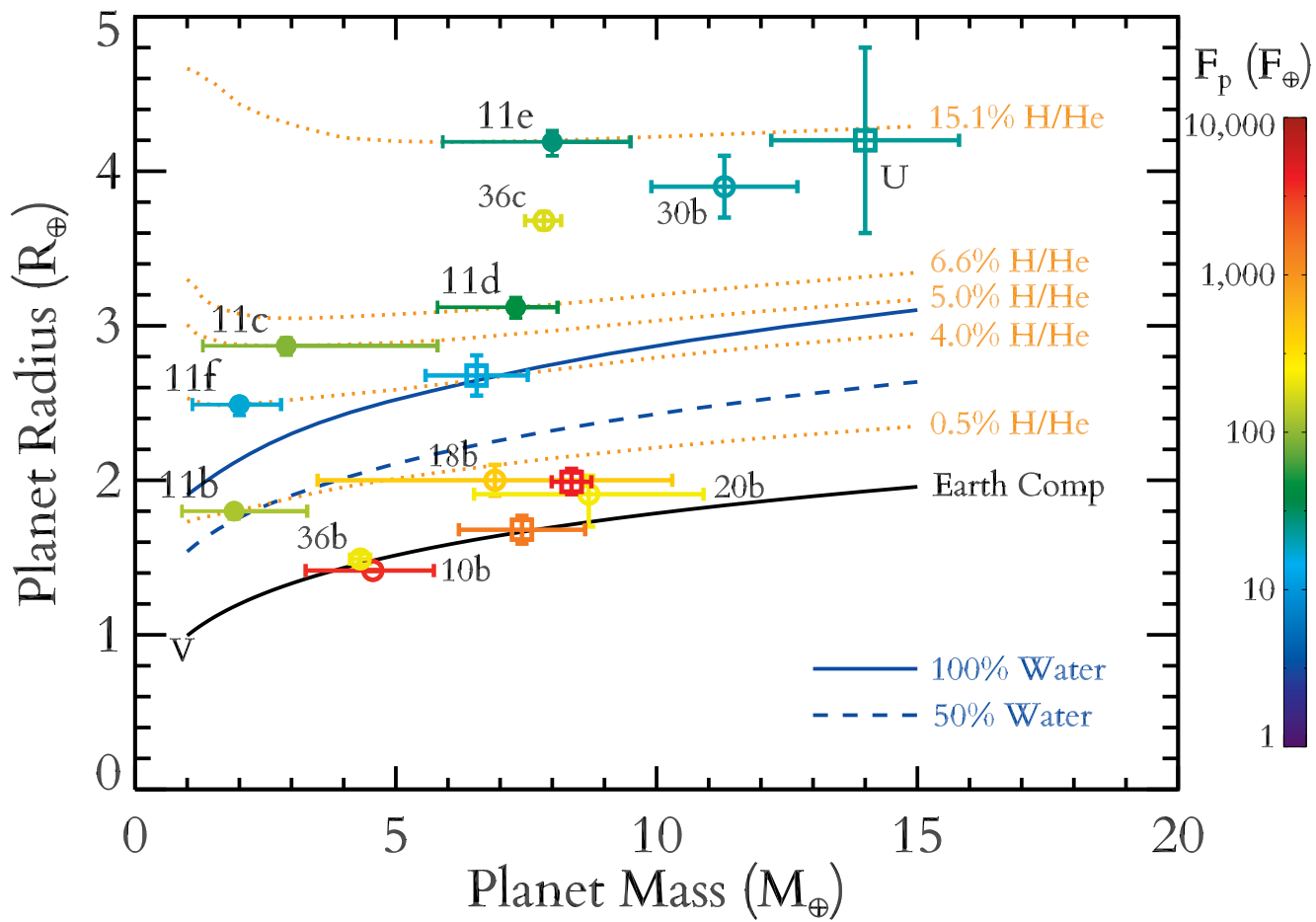}
  \end{center}
  \caption{Updated mass-radius diagram for transiting exoplanets with measured masses, along with curves for different compositions. Planets are color-coded by the incident bolometric flux that they receive. \ik planets are shown by circles, filled for Kepler-11, open for others, with numbers and letters indicating each planet. Other known exoplanets in this mass and radius range are shown by open squares; in order of increasing radius, these are CoRoT-7 b, 55 Cancre e, GJ 1214 b and GJ 3470 b. Solar System planets Venus and Uranus  are shown by black letters. The solid black curve is for an Earth-like composition with 2/3 rock and 1/3 iron by mass. All other curves use thermal evolution calculations \citep{lopez12},
 assuming a volatile envelope atop a  core of rock and iron with composition the same as that of the bulk Earth. The dashed blue curve is for 50\% water by mass,  and the solid blue curve is for a pure H$_2$O planet. The dotted orange curves are for H/He envelopes at 8 Gyr; each one is tailored to match a Kepler-11 planet and is computed at the appropriate flux for that planet.\label{mrfig}}
\end{figure*}

\begin{figure*}[h!] 
  \begin{center}
    \includegraphics[width=7.0in,height=5.0in]{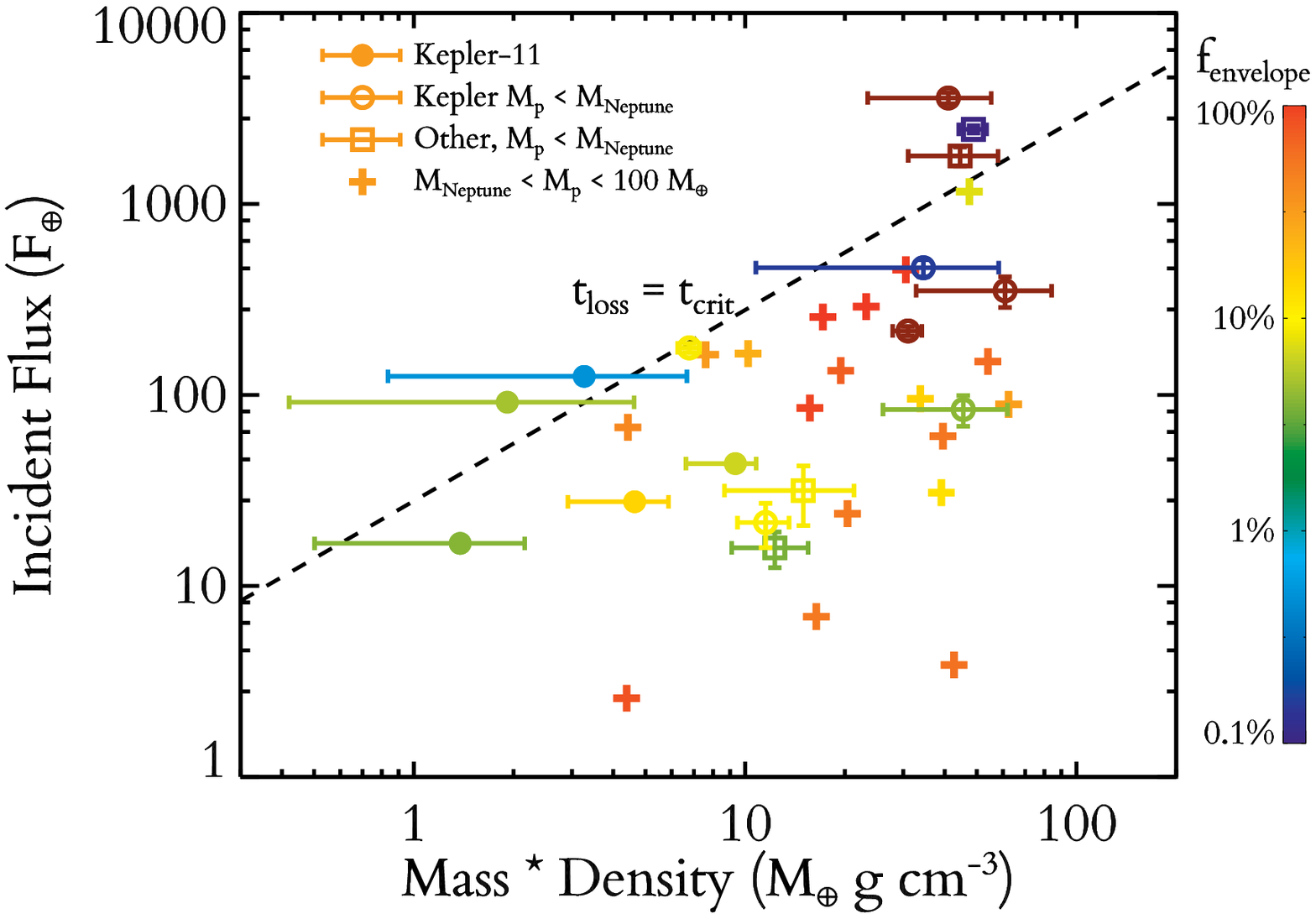}
  \end{center}
  \caption{Updated version of the mass loss threshold diagram from \citet{lopez12}.
 Bolometric flux at the top of the atmosphere, relative to the flux incident on Earth, is plotted against the product of planet mass and planet density. Again, the Kepler-11 planets are shown by filled circles. Open squares show the other extrasolar planets $< 15~M_{\mathrm{\oplus}}$, while crosses show all other transiting planets with measured masses up to 100 $M_{\mathrm{\oplus}}$. Planets are color coded by the percentage of their mass in their H/He envelopes, $f_{envelope}$, according to thermal evolution models. Potentially rocky planets are rust colored. The dashed black line shows the critical mass loss timescale found by \citet{lopez12}.
 \label{thresholdfig}}
\end{figure*}

Combining our dynamical results (as presented in Table~\ref{tbl-dyn} plus upper bounds on the mass of Kepler-11 g illustrated in Figure~\ref{fig:mass-g})  with transit parameters of all planets given in Table~\ref{tbl-transit}, bounds on planet g's eccentricity from transit models, and the stellar characteristics listed in Table~\ref{tbl-star}, we derive the planetary parameters shown in Table~4. The nominal mass values of planets Kepler-11 d, e and f derived herein are within $1 \sigma$ error bars of the preferred fit presented by \citet{liss11a}, and the newly-estimated masses of Kepler-11 b and c are within $2 \sigma$ of their values; the various fits presented by \citet{mig12} are of comparable accuracy. The major differences from the results presented by \citet{liss11a} are that the planetary radii  are $\sim 10\%$ smaller than previously estimated, and planets Kepler-11 b and especially c  are less massive than estimates computed with Q1-Q6 data, resulting in the nominal masses monotonically increasing with planetary radii rather than the inner pair of planets being more dense than the outer ones. Despite the reductions in size estimates, all planets are large for their masses in the sense that they lie above both the $M_p/M_\oplus \approx (R_p/R_\oplus)^{2.06} $ relationship that is valid for planets in our Solar System \citep{liss11b} 
and mass--radius fits to exoplanets \citep{wu12, weis13}.

The six planets in Kepler-11 are all substantially less dense than an iron-free rocky planet, a characteristic already noted for the five inner planets by \citet{liss11a} and \citet{lopez12}, and which now can be stated with even greater (statistical) significance. 
 As a result, they must have substantial envelopes of light components, most likely dominated by the cosmically-abundant constituents H$_2$, He, and/or H$_2$O.  In order to understand these envelopes, we use the thermal evolution models described in detail in \citet{lopez12}. This allows us to determine the size of the H/He envelope for each planet, assuming an Earth-like rock/iron core. 

Figure \ref{mrfig} plots an updated version of the mass--radius diagrams shown in \citet{liss11a} 
and \citet{lopez12}.
 We include all transiting planets with measured masses $M_p < 15~M_{\mathrm{\oplus}}$. 
 For comparison, we include mass--radius curves for Earth-like, 50\% water, and 100\% water compositions. In addition, for each of the five Kepler-11 planets whose mass has been measured, we include a mass--radius curve at the composition (H/He envelope mass fraction) and incident flux of that planet. 
 
The new masses imply that Kepler-11 c is less massive than if it were composed of pure water, meaning that it must have a large H/He envelope. However, Kepler-11 b can still be explained by either a H/He or a steam envelope on top of a rocky core. If we assume that Kepler-11 b's envelope is water rather than H/He, then this planet would be $59\%\pm^{39\%}_{30\%}$ water by mass. The envelope would be composed of steam, since planets like Kepler-11 b are far too irradiated for their interiors to include liquid or high-pressure ice phases. Most of the H$_2$O would be in the vapor and molecular fluid phases, with the ionic fluid and plasma phases occurring at  high pressures deep within these planets (\citealt{nettel08,nettel11}).

For mixtures of rock with H/He (no H$_2$O), and using the sizes and masses presented in Table~4, we find that Kepler-11 b is currently $0.5\%\pm^{0.5\%}_{0.4\%}$ H/He, Kepler-11 c is $5.0\%\pm^{1.1\%}_{0.8\%}$ H/He, Kepler-11 d is $6.6\%\pm^{1.3\%}_{1.2\%}$ H/He, Kepler-11 e is $15.7\%\pm^{1.7\%}_{1.7\%}$ H/He, and Kepler-11 f is $4.0\%\pm^{1.0\%}_{0.7\%}$ H/He by mass. The quoted uncertainties include the measured uncertainties on each planet's mass, radius, incident flux, and age as well as theoretical uncertainties on the albedo and the iron fraction of the rocky/iron core \citep{marcus10}. Despite the small mass fractions in light gases, the presence of these H/He envelopes is key to the observed radii. One way to emphasize this fact is to compare each planet's radius to the radius of its rocky core, as determined by our thermal evolution models for planets lacking H$_2$O. For every  Kepler-11 planet whose mass has been measured except for b, approximately half of the observed radius is due to its H/He envelope. The cores make up 46\%, 54\%, 40\%, and 48\% of the total radii of planets Kepler-11 c, d, e, \& f, respectively, and thus only 6 -- 16\% of the volume. Moreover, even for Kepler-11 b, the rocky core only makes up 66\% of the total radius, corresponding to 29\% of this planet's volume.

In addition, we have included a updated version of the mass loss threshold diagram presented in \citet{lopez12}.
 Figure \ref{thresholdfig} plots incident flux against the product of planet mass times planet density. Diagonal lines  (i.e., lines with slope = 1) in this space correspond to constant mass loss timescales for a specified mean molecular weight of escaping gas, making this diagram useful for understanding how the population of highly-irradiated planets has been sculpted by photoevaporation \citep{lecav07}.
 Here we have color-coded planets by the fraction of their mass in the H/He envelope, assuming an Earth-like core. Four known exoplanets are dense enough to be composed of bare rock (this list includes Kepler-20 b, whose large error ellipse in the mass-radius plane is mostly outside of the rocky composition zone); these planets are shown as rust colored. The key feature of Figure \ref{thresholdfig} is that there is a critical mass loss timescale above which there are no planets with significant H/He envelopes. The dashed black line shows the critical mass loss timescale found by \citet{lopez12}.
 The existence of such a mass loss threshold is a robust predication of planet evolution models that include photoevaporation (\citealt{owen12,lopez12}). The three planets that lie above this threshold in the upper right are Kepler-10 b \citep{bat11}, CoRoT-7 b (\citealt{leger09,que09,hat11}),
 and  55 Cancri e (\citealt{win11,dem11}), none of which are expected to have H/He envelopes.

With the newly-estimated  masses, Kepler-11 b and c  are clearly highly vulnerable to photoevaporation; in fact they both lie on the critical mass loss timescale identified by \citet{lopez12}. On the other hand, planets Kepler-11 d and e have predicted mass loss rates a factor of a few below this threshold.  However, this does not mean that these planets have not experienced significant mass loss. Using the original discovery masses, \citet{lopez12} showed that planets Kepler-11 d and e could have lost at least half of their initial H/He envelopes. Moreover, the assumption of a single critical mass loss timescale is only a rough approximation. The efficiency of photoevaporative mass loss changes as a function of irradiation and stellar age \citep{owen12}. In particular, more irradiated planets like Kepler-11 b and c lose more energy to radiation and recombination-driven cooling, resulting in lower mass loss efficiencies and thus  a higher threshold in Figure 10 \citep{mur09}. This is one possible explanation for why the planets in Kepler-11 do not lie along a single mass-loss timescale.

\section{Conclusions}
We have performed an updated analysis of the Kepler-11 planetary system, concentrating on the dynamical interactions evident in transit timing variations observed in the first 40 months of \ik photometric data.  We have also improved our estimates of the characteristics of the star by combining stellar density constraints from transit profiles and dynamical measurements of planetary eccentricity with spectral information obtained at the Keck observatory.  Our updated  transit, stellar and planetary parameters are presented in Tables~ \ref{tbl-transit}, \ref{tbl-star} 
and 4, respectively. 

The six planets observed to transit Kepler-11 all have small  orbital eccentricities.  None is dense enough to be composed entirely of rocky material, and at least the four middle planets must contain volumetrically-significant envelopes of gases less dense than H$_2$O.  Planets Kepler-11 b and f, and nominally c as well, are less massive than any other exoplanets for which both mass and radius have been measured.  The planetary parameters are consistent with a monotonic increase in mass as a function of radius, although as Figure 9 illustrates, the Kepler-11 planets are less massive for a given radius than most other planets with mass and radius measurements.

\acknowledgments 
{\it Kepler} was competitively selected as the tenth Discovery mission. Funding for this mission is 
provided by NASA's Science Mission Directorate.  E.~A.'s work was supported by NSF Career grant AST-0645416. W.~F.~W.~gratefully acknowledges support from the \ik Participating 
Scientist Program via NASA grant NNX12AD23G, and from the NSF via
grant AST-1109928. D.~J. gratefully acknowledges a Fellowship from the NASA Postdoctoral Program.
We thank Jerome Orosz and Gur Windmiller for assistance in developing
D.S.'s method for measuring transit times
and Tony Dobrovolskis, Darin Ragozzine and Billy Quarles for helpful comments on the manuscript.

\section{ Appendix A: Techniques used to Measure Transit Times}
We measured transit times using three different techniques, each of which is described below.  
\subsection{TT Measurements by Jason Rowe}
This analysis used Q1 -- Q14 long cadence and Q3 -- Q14 short cadence \ik simple aperture photometry (labeled SAP\_FLUX).  Only  data with a quality flag set to zero as documented in the \ik data release notes were used.  This provided 52,539 and 1,464,980 
long and short cadence photometric measurements, respectively.  

The data were initially detrended using a running 2-day box-car median filter that was applied to individual segments of time-series photometry.  A segment was defined as a continuous  string of time-series data that does not contain an interruption longer than 2.5 hours (5 long cadence measurements).  This was done to handle offsets observed after data outages, typically caused by a change in the thermal environment of the CCD detector.  A circular quadratic transit model based on \citet{man02} was fit to the data by minimization of $\chi^2$ with a Levenberg-Marquardt  algorithm.  The transit model was used to measure the transit duration for each transiting planet.  The original SAP\_FLUX photometric data were then reprocessed using a second-order polynomial to detrend the time-series to remove instrumental (such as focus changes) and astrophysical effects.  All data obtained during transit were excluded, as well as those taken in the 30 minutes before ingress and in the 30 minutes after egress.  A clipping algorithm was used to exclude any measurement that differed from the mean by more than $3\sigma$.  Measurements obtained during a planet transit were excluded from the clipping exercise.  It was found that the data before a data outage near JD = 2455593 could not be sufficiently detrended.  As such, data from 2455593 to 2455594.5 were excluded, which meant that a transit of Kepler-11 g was not included in our analysis.

The detrended LC and SC photometric time-series were then each fit with a multi-planet, circular orbit, quadratic Mandel \& Agol transit model.  The model parameters are the mean stellar density (\rhostar), photometric zero point, and, for each planet, the center of transit time, orbital period, impact parameter, and scaled planetary radius (\rprs).  The model assumes that the mass of star is much greater than the combined mass of the orbiting planets, so that
\begin{equation}\label{eq:rhostar}
\left(\frac{a}{R_\star}\right)^3 \frac{3\pi}{G P^2} = \frac{(M_\star+M_p)}{\frac{4 \pi}{3}R_\star^3} \approx \rhostar.
\end{equation}

A photometric time-series for each transiting planet was then produced by removing the transits of the other transiting planets.  The remaining transits were then individually fit by using the best-fit model as a template and only allowing the center of transit time to vary.  This yielded a time-series of transit timing variations (TTVs) for each planet.  The measured TTVs were then used to linearize (or {\it deTTV}) the photometry, such that when folded at the orbital period the transits are aligned in the resulting lightcurve.  The multi-planet transit model was then refit out to the {\it deTTVed} lightcurve and used the updated template to determine the final set of TTVs shown by the green points in Figure 1.  Uncertainties in the transit times were determined by examining the residuals from the fits to each individual transits and scaling the photometric errors such that reduced $\chi^2$  was equal to one.   The diagonal elements of the co-variance matrix were adopted as the uncertainty in the measurement.

\subsection{TT Measurements by Eric Agol} 
The times of transit were fit using a quadratic limb-darkening model in 
which the duration and impact parameter for each planet were assumed to 
be fixed, while the times of each transit were allowed to vary.  The 
model was computed simultaneously for the short cadence (when available) 
and long cadence data (otherwise).  A window equal to one transit 
duration was included before and after every transit.  The lightcurve 
was divided by the model (computed for all planets simultaneously so 
that overlapping transits were properly accounted for), and then fit 
with a third-order polynomial for each contiguous data set (without gaps 
larger than 12 hours).  The model parameters were optimized until a best 
fit was found;  a second iteration was carried out after outliers from 
the first fit were rejected.
After finding the best fit, the times of each and every transit were 
allowed to vary over a grid of values spanning (typically) about 2 hours 
on either side of the best fit time.  The variation in $\chi^2$ with 
transit time was then fit with a quadratic function to measure the 
uncertainty in the transit time.  If that fit failed, then the transit 
time error was measured from the width of the $\chi^2$ function for 
values less than one above the best fit value.

\subsection{TT Measurements by Donald Short} 
In contrast to the Rowe and Agol methods, a purely mathematical 
technique was used to determine the transit times, under the assertion 
that the time of a transit event can be estimated without need of a 
physical model of the event. Under conditions of poor signal-to-noise 
ratio or undersampling, the constraints imposed by a physical model 
are extremely valuable. For high signal-to-noise cases, a non-physical 
model can match, or even excel a physical model under certain 
conditions. The limitations in a physical model, such as imperfect 
limb darkening parameterization or assumed zero eccentricity, have 
no consequence in a non-physical model. Since no assumptions about 
sphericity, obliquity, gravity darkening, strict keplerian motion, 
etc., were made, the method is insensitive to errors in these 
physical parameters or effects.

Both LC and SC data were used in computing the planetary 
transit time estimates, provided the pipeline data quality flag had 
the nominal value of zero. The TTs were estimated by an 
iterative method starting with the SC data. Using an 
estimate of the transit duration and estimates of the transit times 
based on the linear ephemeris from \citet{liss11a},
each transit was locally detrended. Detrending employed a low-order 
polynomial centered on the transit and extending symmetrically either 
0.3, 0.6, or 0.83 days beyond the ends of the transit; the length and 
polynomial order that provided the best fit to these out-of-transit 
data was selected. During this process, each transit was checked for 
missing data and overlapping transits from other planets that could 
compromise the determination of that TT. Transits that had such 
problems were eliminated from further consideration.  After 
detrending, the transits were shifted in time so that the center 
of each transit was at time zero. All of the transits were then 
combined (``stacked'' or ``folded'' on top of each other). A 
piecewise cubic  Hermite spline (PCHS) was then least-squares fit 
to the combined-transit lightcurve, giving a transit template. 
The transit template was generated by the data themselves; no 
physical constraints on its shape were imposed. As such, it should 
be an excellent match to the observed transits. From this template, 
a refined transit width was estimated and used to revise the 
detrending of each transit. The template was then correlated with 
each individual transit, yielding improved TTs. Any outliers with 
respect to the template were flagged and eliminated from further 
template building, but no rejections were made when estimating the 
individual TTs. The detrended transits were shifted (folded) on the 
revised TTs, combined, and a new PCHS template generated. Again, the 
individual transits were then detrended, now using both the revised 
duration and revised transit times. The detrended transits were 
correlated with the revised template, yielding a refined set 
of TTs. Three iterations of this process were carried out. The 
uncertainty in each TT was estimated from the shifts in time needed 
to degrade the $\chi^2$ fit of the template to the transit by one.

For transits with LC data only, the SC PCHS 
template was convolved to 30 minutes, yielding the LC 
template. The LC template was then correlated with each 
transit, providing a correction to the times from the initial linear 
ephemeris. The revised TTs were used to improve the detrending window, 
but the template was not updated-- it was held fixed at the shape 
derived from the SC template. This process iteratively 
produced measurements of the TTs, uncertainties, and model fits for 
each transit. Finally, those TTs that had large timing error 
estimates ($>$40 minutes) were eliminated from the final list of TTs.

The process above was repeated independently for each planet, noting 
that overlapping transits from different planets were discarded. In 
general, the TTs computed by this method agree quite well with the physical methods; however, the error estimates are notably larger.

\section{ Appendix B: Details of Dynamical Models}
Here we present the results of our dynamical models in detail. We carried out three classes of fit using each set of TTs.  In the ``all-circular'' class, all planets were assumed to travel on circular orbits at epoch.  In the ``all-eccentric'' class, all planets were allowed to have eccentric orbits at epoch.  We found that the quality of these fits was not sensitive to the mass or eccentricity of planet Kepler-11 g as long as these were not too large, so we performed ``g-fixed'' fits  wherein the eccentricity of  planet g is set to zero at epoch and its mass set to $25.3\times10^{-6}~M_{\star}$ (which equals $8~M_{\oplus}$ for an assumed stellar mass of $0.95~M_{\oplus}$, as estimated by \citealt{liss11a}).  

Table \ref{tbl-fits} compares the quality of fit between using various data sets and assumptions.  Note that comparisons of the numerical values between the quality of fits using different data sets are not meaningful because of the differing prescriptions employed to compute the uncertainties of individual TTs, but comparison between the reduced $\chi^2$ for the all-circular, all-eccentric, and g-fixed results using a given set of TTs shows that eccentricities are detected for the five inner planets but not for planet g.  As the quality of the all-circular fits are distinctly inferior to those that allow at least the five inner planets to travel on eccentric orbits, we do not consider the all-circular fits further.

\begin{table}
  \begin{center}
    \begin{tabular}{|c|ccc|ccc|ccc|}
      \hline
Planet  & All          & circular   &    &  All        & eccentric  &     & g-fixed       & &  (Tables \ref{tbl-EAbestfit},~\ref{tbl-JRbestfit},~\ref{tbl-DSbestfit})  \\
        & $\chi^2_{EA}$ & $\chi^2_{JR}$  &   $\chi^2_{DS}$  &    $\chi^2_{EA}$ & $\chi^2_{JR}$  &  $\chi^2_{DS}$  &   $\chi^2_{EA}$  &  $\chi^2_{JR}$  & $\chi^2_{DS}$   \\
      \hline
              b & 229.62  &  100.36  & 41.61      & 189.49  &  86.59   & 36.16      & 189.43 & 86.61  & 36.14 \\
              c & 280.39  &  111.72  & 72.30      &  211.18 &  79.71   & 50.15      & 211.21 & 79.66  & 50.14 \\
              d & 123.25  &  51.68   & 15.63      &  100.95 &  45.08   &  14.78      & 101.02 & 45.08  & 14.79 \\
              e & 78.47   &  53.97   & 22.06      &  46.47  &  29.08   &  12.03      & 46.73  & 29.24  & 12.06 \\ 
              f & 120.79  &  58.09   & 38.70      &  52.03  &  16.81   &  12.73      & 54.10  & 17.24  & 13.06 \\
              g & 9.84    &  6.71    & 3.69       &  10.08  &  6.69    &  3.62       & 9.61   & 6.66   & 3.64  \\
      \hline
           total & 842.16  & 382.53  & 193.99      & 610.20  & 263.97   & 129.46       & 612.09 & 264.49  & 129.83 \\
 $\chi^2/$(d.o.f.) & 3.25    & 1.38    & 0.82        & 2.47    &  1.00   & 0.58        & 2.46  & 0.99  &  0.57  \\
       \hline
    \end{tabular}
    \caption{$\chi^2$ contributions from each planet for a suite of models against both sets of transit times. The second through fourth columns show best fits to an orbital configuration with all eccentricities fixed at zero, the fifth through seventh columns show all eccentric fits, and the eighth through tenth columns shows the g-fixed models whose results are shown in Tables \ref{tbl-EAbestfit}, \ref{tbl-JRbestfit}} and \ref{tbl-DSbestfit}.
\label{tbl-fits}
  \end{center}
\end{table}

As shown in Table \ref{tbl-fits}, the g-fixed fits, which are presented in Tables \ref{tbl-EAbestfit}, \ref{tbl-JRbestfit} and \ref{tbl-DSbestfit},  are of slightly better quality (in a $\chi^2/$d.o.f. sense) than are the corresponding all-eccentric fits. Thus, the parameters from the three g-fixed fits are synthesized to incorporate the full ranges of all $1\sigma$ error bars from fits to each set of data and displayed as our primary results in Table~\ref{tbl-dyn}. Table \ref{tbl-allecc} is the counterpart of Table 1, synthesizing all-eccentric fit results of the three sets of transit time data.

The small values (compared to unity) of $\chi^2/$(d.o.f.)  shown in Table~\ref{tbl-fits} for fits to D.S.'s TTs imply that the uncertainties quoted for these TTs were overestimated.  Similarly, the large values of $\chi^2/$(d.o.f.)  for E.A.'s TTs strongly suggest that these uncertainties were underestimated.  The values of $\chi^2/$(d.o.f.) near unity for both fits allowing eccentric planetary orbits to J.R.'s TTs suggest that uncertainties in these TTs may have been slightly overestimated.

\begin{table}
  \begin{center}
    \begin{tabular}{|cccccc|}
      \hline
      \hline
      Planet  &  $P$ (days)  & $T_{0}$ &   $e \cos \omega$ & $e \sin \omega$   & $M_{p}/M_{\star} \times 10^{-6}$ \\
\hline 
b  & \textbf{10.3043}$\pm 0.0002$ & \textbf{689.7378}$\pm 0.0009 $ & $0.038\pm 0.016$ & $0.009 \pm 0.008 $ & \textbf{3.91}$ \pm 1.03 $ \\  
 c  & \textbf{13.0236}$\pm 0.0003$ & \textbf{683.3494}$\pm 0.0005 $ & $0.019\pm 0.014$ & $-0.005 \pm 0.004 $ & \textbf{6.23}$ \pm 1.75 $ \\  
 d  & \textbf{22.6839}$\pm 0.0003$ & \textbf{694.0061}$\pm 0.0005 $ & $-0.006\pm 0.003$ & $0.001 \pm 0.001 $ & \textbf{23.60}$ \pm 1.66 $ \\  
 e  & \textbf{31.9996}$\pm 0.0004$ & \textbf{695.0752}$\pm 0.0005 $ & $-0.009\pm 0.002$ & $-0.009 \pm 0.001 $ & \textbf{27.77}$ \pm 1.92 $ \\  
 f  & \textbf{46.6903}$\pm 0.0011$ & \textbf{718.2737}$\pm 0.0015 $ & $0.007\pm 0.003$ & $-0.007 \pm 0.002 $ & \textbf{7.45}$ \pm 1.09 $ \\  
 g  & \textbf{118.3807}$\pm 0.0004$ & \textbf{693.8022}$\pm 0.0010 $ &  (0) & (0) & (25.29) \\  
      \hline
        \hline
    \end{tabular}
    \caption{Best dynamical fit (fixed mass and circular orbit for planet g) to TTs from E.A. \label{tbl-EAbestfit}}
  \end{center}
\end{table}

\begin{table}
  \begin{center}
    \begin{tabular}{|cccccc|}
      \hline
      \hline
        Planet  &  $P$ (days)  & $T_{0}$ &   $e \cos \omega$ & $e \sin \omega$   & $M_{p}/M_{\star} \times 10^{-6}$ \\
    \hline
b   & \textbf{10.3039}$\pm 0.0004$ & \textbf{689.7391}$\pm 0.0012 $ & $0.050\pm 0.019$ & $0.014 \pm 0.010 $ & \textbf{6.80}$ \pm 2.16 $ \\  
c   & \textbf{13.0240}$\pm 0.0005$ & \textbf{683.3497}$\pm 0.0010 $ & $0.033\pm 0.016$ & $0.005 \pm 0.008 $ & \textbf{9.25}$ \pm 3.34 $ \\  
d   & \textbf{22.6849}$\pm 0.0005$ & \textbf{694.0072}$\pm 0.0007 $ & $-0.003\pm 0.004$ & $0.004 \pm 0.003 $ & \textbf{23.61}$ \pm 1.84 $ \\  
e   & \textbf{31.9999}$\pm 0.0005$ & \textbf{695.0756}$\pm 0.0005 $ & $-0.007\pm 0.003$ & $-0.007 \pm 0.003 $ & \textbf{23.85}$ \pm 2.55 $ \\  
f   & \textbf{46.6877}$\pm 0.0014$ & \textbf{718.2697}$\pm 0.0021 $ & $0.014\pm 0.005$ & $-0.001 \pm 0.002 $ & \textbf{5.26}$ \pm 1.21 $ \\  
g   & \textbf{118.3806}$\pm 0.0005$ & \textbf{693.8010}$\pm 0.0010 $ & (0) & (0) & (25.29) \\
      \hline
        \hline
    \end{tabular}
    \caption{Best (g-fixed) dynamical fit to TTs from J.R. \label{tbl-JRbestfit}}
  \end{center}
\end{table}

\begin{table}
  \begin{center}
    \begin{tabular}{|cccccc|}
      \hline
      \hline
      Planet  &  $P$ (days)  & $T_{0}$ &   $e \cos \omega$ & $e \sin \omega$   & $M_{p}/M_{\star} \times 10^{-6}$ \\
\hline
b   & \textbf{10.3036}$\pm 0.0007$ & \textbf{689.7363}$\pm 0.0032 $ & $0.009\pm 0.008$ & $0.072 \pm 0.018 $ & \textbf{6.80}$ \pm 3.28 $ \\  
c   & \textbf{13.0247}$\pm 0.0006$ & \textbf{683.3490}$\pm 0.0015 $ & $-0.004\pm 0.005$ & $0.059 \pm 0.014 $ & \textbf{12.06}$ \pm 6.25 $ \\  
d   & \textbf{22.6846}$\pm 0.0006$ & \textbf{694.0074}$\pm 0.0016 $ & $-0.001\pm 0.002$ & $-0.000 \pm 0.001 $ & \textbf{21.27}$ \pm 3.23 $ \\  
e   & \textbf{31.9993}$\pm 0.0009$ & \textbf{695.0759}$\pm 0.0011 $ & $-0.008\pm 0.003$ & $-0.010 \pm 0.004 $ & \textbf{22.91}$ \pm 4.73 $ \\  
f   & \textbf{46.6883}$\pm 0.0027$ & \textbf{718.2695}$\pm 0.0023 $ & $0.014\pm 0.007$ & $-0.007 \pm 0.005 $ & \textbf{5.94}$ \pm 2.55 $ \\  
g   & \textbf{118.3809}$\pm 0.0003$ & \textbf{693.8030}$\pm 0.0021 $ &  (0) & (0) & (25.29)  \\  
      \hline
        \hline
    \end{tabular}
    \caption{Best (g-fixed) dynamical fit to TTs from D.S. \label{tbl-DSbestfit}}
  \end{center}
\end{table}

\begin{table}
  \begin{center}
    \begin{tabular}{|cccccc|}
      \hline
      \hline
     Planet  &  $P$ (days)  & $T_{0}$ &   $e \cos \omega$ & $e \sin \omega$   & $M_{p}/M_{\star} \times 10^{-6}$ \\
\hline
b   & \textbf{10.3039}$^{+0.0006}_{-0.0011}$ & \textbf{689.7377}$^{+0.0031}_{-0.0046}$ & $0.032^{+0.037}_{-0.035}$ & $0.032^{+0.060}_{-0.030}$ & \textbf{5.83}$^{+4.29}_{-3.09}$ \\  
c   & \textbf{13.0241}$^{+0.0013}_{-0.0008}$ & \textbf{683.3494}$^{+0.0014}_{-0.0020}$ & $0.016^{+0.035}_{-0.029}$ & $0.020^{+0.054}_{-0.030}$ & \textbf{9.13}$^{+9.30}_{-4.77}$ \\  
d   & \textbf{22.6845}$^{+0.0010}_{-0.0009}$ & \textbf{694.0069}$^{+0.0022}_{-0.0013}$ & $-0.003^{+0.006}_{-0.006}$ & $0.002^{+0.006}_{-0.002}$ & \textbf{22.84}$^{+2.64}_{-4.97}$ \\  
e   & \textbf{31.9996}$^{+0.0008}_{-0.0013}$ & \textbf{695.0755}$^{+0.0015}_{-0.0008}$ & $-0.008^{+0.005}_{-0.004}$ & $-0.009^{+0.004}_{-0.005}$ & \textbf{24.83}$^{+4.84}_{-7.05}$ \\  
f   & \textbf{46.6887}$^{+0.0029}_{-0.0038}$ & \textbf{718.2711}$^{+0.0043}_{-0.0052}$ & $0.011^{+0.010}_{-0.007}$ & $-0.005^{+0.006}_{-0.007}$ & \textbf{6.20}$^{+2.52}_{-2.93}$ \\  
g   & \textbf{118.3809}$^{+0.0012}_{-0.0010}$ & \textbf{693.8021}$^{+0.0030}_{-0.0022}$ & $0.032^{+0.097}_{-0.103}$ & $0.022^{+0.055}_{-0.063}$ & \textbf{23.21}$^{+59.18}_{-58.69}$ \\  
      \hline
        \hline
    \end{tabular}
    \caption{Dynamical all-eccentric fits to the observed transit times with the orbital periods (second column), time of first transit after epoch (third column), $e\cos\omega$ (fourth column; $e$ represents eccentricity, $\omega$ is the angle, measured from the star, between the place the planet's orbit pierces the sky, coming towards the observer, and  the pericenter of the orbit), $e\sin\omega$ (fifth column), and planetary mass in units of the stellar mass (sixth column), all as free variables. For planet g, this model has settled on a mass near the initial estimate of 8 $M_{\oplus}$ ($25.3\times10^{-6}~M_{\star}$). \label{tbl-allecc}}
  \end{center}
\end{table}

\end{document}